\documentclass[aps,10pt,prd,notitlepage,showpacs,nofootinbib,superscriptaddress,compressed]{revtex4-1}

\pdfoutput=1 
\usepackage[utf8]{inputenc}
\usepackage[english]{babel}
\usepackage{amsmath}
\usepackage{graphicx}
\usepackage{dcolumn}
\usepackage{pbox}
\usepackage{amssymb}
\usepackage{epsfig}
\usepackage[dvipsnames]{xcolor}
\usepackage{slashed}
\usepackage{amssymb}
\usepackage{ mathrsfs }
\usepackage{color}
\usepackage[font=small]{caption}
\usepackage[font=small]{subcaption}
\usepackage{url}
\usepackage{braket}
\usepackage{hyperref}
\usepackage{fontawesome5}
\usepackage{booktabs}
\graphicspath{{pic/}}

\definecolor{DarkBlue}{rgb}{0.7, 0.4, 1} 
\definecolor{Blue}{rgb}{0, 0.8, 0} 
\definecolor{MyLightBlue}{rgb}{0.5,0.7,1.9}
\definecolor{MyGreen}{rgb}{0.0,0.2, 0.0}
\definecolor{MyBrickRed}{rgb}{0, 0.5, 0.2}
\RequirePackage{hyperref}
\hypersetup{colorlinks, citecolor=Blue,linkcolor=DarkBlue, urlcolor=Green}

\newcommand{\bea}{\begin{eqnarray}}
\newcommand{\eea}{\end{eqnarray}}
\newcommand\dd{\mathrm{d}}

\makeatletter

\makeatletter
\renewcommand\@makecaption[2]{%
  \par
  \vskip\abovecaptionskip
  \begingroup
  
   \small\rmfamily
    \begingroup
     \samepage
     \flushing
     \let\footnote\@footnotemark@gobble
     \@make@capt@title{#1}{#2}\par
    \endgroup
  \endgroup
  \vskip\belowcaptionskip
}
\makeatother

\DeclareUnicodeCharacter{2212}{-}
\newcommand {\black} {\color{black}}

\def\21{\mathrm{$SU(2)_L \otimes U(1)_Y$}}

\begin{document}
\title{Forward Searches for Heavy Neutrinos and $Z'$ Bosons at FCC-hh}
\author{ShivaSankar K.A.}\email{a-shiva@particle.sci.hokudai.ac.jp}
\affiliation{Department of Physics, Hokkaido University, Sapporo 060-0810, Japan}
\author{Souvik Das}\email{souvik\_das@tamu.edu}
\affiliation{Department of Physics and Astronomy, Mitchell Institute for Fundamental Physics and Astronomy, Texas A\&M University, College Station, Texas 77843, USA}
\author{Arindam Das}
\email{adas@particle.sci.hokudai.ac.jp}
\affiliation{Institute for the Advancement of Higher Education, Hokkaido University, Sapporo 060-0817, Japan}
\affiliation{Department of Physics, Hokkaido University, Sapporo 060-0810, Japan} 
\author{Sanjoy Mandal}
\email{smandal@kias.re.kr}
\affiliation{Korea Institute for Advanced Study, Seoul 02455, Korea} 
\begin{abstract}
The discovery of neutrino masses strongly motivates extensions of the Standard Model containing heavy neutral leptons and additional gauge interactions. We investigate the prospects for probing these states at the proposed Forward Physics Facility (FPF) of the 100 TeV Future Circular Collider (FCC-hh) within a broad class of anomaly-free chiral $U(1)$ gauge extensions. These models predict a new neutral gauge boson, $Z'$, together with right-handed neutrinos responsible for generating light neutrino masses through the seesaw mechanism. We study long-lived particle signatures arising from both heavy neutrinos and the $Z'$ boson produced in the far-forward region. In particular, we analyze heavy neutrino production from meson decays, visible decays of long-lived $Z'$ bosons produced through meson decays and proton bremsstrahlung, long-lived $Z'$ bosons decaying into heavy-neutrino pairs, and prompt $Z'$ decays yielding long-lived heavy neutrinos. The expected event rates are evaluated for the proposed FPF detector configurations, taking into account realistic detector geometry, decay probabilities, and visible final states. We derive projected sensitivities to the heavy neutrino mass and active-sterile mixing as well as to the $Z'$ mass and gauge coupling for several representative $U(1)$ charge assignments. Our results demonstrate that the FPF at FCC-hh can substantially extend the discovery reach for light long-lived heavy neutrinos and light $Z'$ bosons beyond existing and proposed experiments, providing a powerful and complementary probe of neutrino-mass models and hidden gauge sectors.\href{https://github.com/SouvikPhD/RHN-Detection-with-FASER-2-}{\faGithub}
\end{abstract}
\maketitle
\section{Introduction}
The observation of non-zero neutrino masses and flavor mixing~\cite{ParticleDataGroup:2020ssz} provides compelling evidence for physics beyond the Standard Model~(BSM). Among the simplest and most attractive explanations is the type-I seesaw mechanism~\cite{Minkowski:1977sc,Yanagida:1979as,Gell-Mann:1979vob,Mohapatra:1979ia,Yanagida:1980xy,Schechter:1980gr}, which augments the SM with heavy neutral leptons (HNLs), commonly referred to as right-handed neutrinos (RHNs). As gauge singlets, these heavy states naturally explain the tiny masses of the light neutrinos and their flavor mixing, while providing a window into physics at scales well above the electroweak scale.
\par A particularly well-motivated realization of the seesaw mechanism is obtained by extending the SM gauge symmetry with an additional $U(1)$ group. Such models predict a new neutral gauge boson, $Z^\prime$, together with an SM-singlet scalar whose vacuum expectation value~(VEV) spontaneously breaks the new gauge symmetry. Gauge anomaly cancellation requires the introduction of three RHNs, which simultaneously generate light neutrino masses through the seesaw mechanism. The simplest realization is the $U(1)_{B-L}$ model~\cite{Davidson:1978pm,Davidson:1979wr,Marshak:1979fm,Mohapatra:1980qe}, while more general anomaly-free chiral $U(1)_X$ extensions~\cite{Appelquist:2002mw,Coriano:2014mpa,Das:2017flq} assign different charges to left- and right-handed fermions, leading to chiral $Z^\prime$ interactions. Alternative anomaly-free constructions with non-universal RHN charges have also been proposed~\cite{Das:2017deo}. Since the $Z^\prime$ mass and the $U(1)_X$ gauge coupling remain free parameters, these models give rise to a rich collider phenomenology that can be explored over a broad range of energies~\cite{Das:2021esm,Asai:2022zxw,Asai:2023mzl,KA:2023dyz,A:2024shl}.
\par Because the RHNs carry $U(1)_X$ charge, they can be produced directly through $Z^\prime$ decays whenever kinematically allowed, leading to prompt or displaced signatures at colliders~\cite{Das:2019fee,Chiang:2019ajm,Cvetic:2018elt,Cvetic:2019rms}. For sufficiently light $Z^\prime$ bosons, RHNs can also be produced through meson decays and proton bremsstrahlung, making forward experiments particularly sensitive to these scenarios. Such signatures have been extensively investigated for the FASER experiment~\cite{Feng:2017uoz,Kling:2021fwx}, including RHN pair production in the $B-L$ model~\cite{Li:2023dbs} and RHN production from meson decays in low-scale seesaw scenarios~\cite{Kling:2018wct,Feng:2025adw,A:2025gpb}. Complementary searches are also being pursued at MoEDAL-MAPP~\cite{Frank:2019pgk} and proposed detectors such as FACET~\cite{Cerci:2021nlb}, MATHUSLA~\cite{Chou:2016lxi}, CODEX-b~\cite{Gligorov:2017nwh}, and ANUBIS~\cite{Hirsch:2020klk}. These studies demonstrate the excellent sensitivity of forward long-lived particle~(LLP) searches to the heavy-neutrino mass and active-sterile mixing parameter space~\cite{Chun:2019nwi}.
\par In our previous work~\cite{A:2025ygb}, we investigated RHN pair production via a light $Z^\prime$ boson at the ATLAS interaction point, with detection prospects at the proposed FASER2 experiment. The $Z^\prime$ was assumed to be produced through meson decays and proton bremsstrahlung~\cite{Feng:2017vli,FASER:2018eoc,FASER:2018bac}, and we considered two complementary scenarios: a long-lived $Z^\prime$ decaying visibly inside the detector and a prompt $Z^\prime$ decaying into RHN pairs. We demonstrated that the chiral structure of general $U(1)_X$ models plays a crucial role in determining the $Z^\prime$ decay widths and branching fractions into heavy neutrinos.
\par A natural extension of these studies is the proposed Forward Physics Facility (FPF) at the 100 TeV Future Circular Collider (FCC-hh)~\cite{FCC:2018byv,FCC:2018vvp,Aleksa:2019pvl,MammenAbraham:2024gun}. With a center-of-mass energy of $\sqrt{s}=100$ TeV and an integrated luminosity of up to $30~\mathrm{ab}^{-1}$, the FCC-hh will produce an enormous flux of light particles in the far-forward region, making the FPF an ideal facility to search for feebly interacting particles~(FIPs) and long-lived particles~(LLPs) that are difficult to probe at conventional collider detectors. Located about 1.5 km downstream of the interaction point, the FPF is designed to detect visible decays of LLPs while a dedicated sweeper magnet suppresses the charged-particle background.
Following Ref.~\cite{MammenAbraham:2024gun}, we consider two detector configurations optimized for LLP searches. The first, FPF1, consists of a $5\,\mathrm{m}\times5\,\mathrm{m}\times50\,\mathrm{m}$ decay volume, comparable in size to SHiP~\cite{Ahdida:2704147,SHiP:2025ows} and larger than FASER2, whereas the second, FPF2, employs a much larger $20\,\mathrm{m}\times20\,\mathrm{m}\times400\,\mathrm{m}$ decay volume. Both designs assume a visible-energy threshold of $E_{\rm vis}\ge100$ GeV and excellent charged-particle reconstruction using a magnetic spectrometer. Incoming charged particles are rejected by veto systems, neutrino-induced backgrounds are suppressed by maintaining the decay volume under vacuum, and a downstream electromagnetic calorimeter enhances the reconstruction of electrons and positrons from LLP decays.
\par In this work, we consider a class of anomaly-free chiral $U(1)_X$ extensions of the SM. We first study the minimal realization of Ref.~\cite{Das:2017flq}, which contains three SM-singlet RHNs with universal $U(1)_X$ charges and one SM-singlet scalar. After the spontaneous breaking of the $U(1)_X$ and electroweak symmetries, the RHNs acquire Majorana and Dirac masses, respectively, generating light neutrino masses through the type-I seesaw mechanism. Anomaly cancellation leads to asymmetric $U(1)_X$ charges for left- and right-handed fermions, giving rise to chiral $Z^\prime$ interactions. We also consider the $U(1)_{xq-\tau_R^3}$ and $U(1)_{q+xu}$ realizations~\cite{Carena:2004xs,Hashimoto:2014ela}, which preserve the chiral structure of the $Z^\prime$ interactions while differing in their charge assignments.
In addition, we investigate the alternative anomaly-free $U(1)_X$ model of Ref.~\cite{Das:2017deo}, where two RHNs carry identical $U(1)_X$ charges while the third has a distinct charge. This framework realizes a neutrinophilic two-Higgs-doublet model in which only two RHNs participate in the seesaw mechanism, whereas the third remains protected by the $U(1)_X$ symmetry and naturally serves as a dark matter candidate~\cite{Okada:2018tgy}. In all these scenarios, the seesaw-induced light-heavy neutrino mixing generates interactions of heavy neutrinos with the SM weak gauge bosons, while their non-zero $U(1)_X$ charges allow direct couplings to the $Z^\prime$. Consequently, single heavy neutrino production through charged-current interactions is suppressed by the light-heavy mixing, whereas heavy-neutrino pair production proceeds directly via the $Z^\prime$.

In this paper, we investigate heavy neutrino production at the FPF through two complementary mechanisms. First, we study heavy neutrinos produced in meson decays at the FCC-hh interaction point. Owing to their long lifetimes, these particles can travel to the FPF detectors before decaying into visible final states, allowing us to probe the heavy-neutrino mass and active-sterile mixing. This search strategy is equally applicable to light sterile neutrinos predicted in low-scale seesaw models~\cite{Gorbunov:2007ak,Bolton:2019pcu,Fernandez-Martinez:2023phj}.
We also investigate heavy-neutrino pair production via a $Z^\prime$ produced through meson decays and proton bremsstrahlung. Three complementary scenarios are considered: (i) a long-lived $Z^\prime$ with the $Z^\prime\to NN$ channel kinematically forbidden, leading to visible $Z^\prime$ decays; (ii) a long-lived $Z^\prime$ decaying into long-lived heavy neutrinos, some of which may escape the detector before decaying; and (iii) a prompt $Z^\prime$ decaying into long-lived heavy-neutrino pairs. By analyzing the visible decays of both the $Z^\prime$ and heavy neutrinos, we derive projected sensitivities to the $U(1)_X$ gauge coupling, the $Z^\prime$ mass, the heavy-neutrino mass, and the active-sterile mixing for several representative $U(1)$ charge assignments.
\par This paper is organized as follows. In Sec.~\ref{sec:model}, we introduce the general $U(1)$ models and summarize the relevant interactions and decay widths of the $Z^\prime$ boson and heavy neutrinos. In Sec.~\ref{sec:experiment}, we describe the FPF@FCC setup, the signal analyses, and the projected sensitivities for the different long-lived particle scenarios. We conclude in Sec.~\ref{sec:conc}.
\section{General $U(1)$ extensions of the Standard Model}
\label{sec:model}
We consider two classes of anomaly-free $U(1)$ extensions of the SM. The first is a minimal framework containing three generations of SM-singlet RHNs and one SM-singlet scalar, where all three RHNs carry identical $U(1)$ charges. The second extends the scalar sector and features non-universal $U(1)$ charge assignments for the RHNs. These two scenarios are described below.
\subsection{Case-I}
A minimal $U(1)$ extension of the SM consists of three generations of SM-singlet RHNs together with an SM-singlet scalar field $\Phi$. Since the RHNs are charged under the additional $U(1)$ gauge symmetry, they play a crucial role in cancelling both gauge and mixed gauge-gravitational anomalies. The complete particle content together with the corresponding general $U(1)$ charge assignments, obtained after imposing the anomaly cancellation conditions, is summarized in Table~\ref{tab1}.
\begin{table}[h]
	\begin{center}
		\begin{tabular}{| c| c || c | c |c||c|}
			\hline
			\hspace{0.5cm}Fields \hspace{0.5cm}   & \hspace{0.5cm} $SU(3)_c\otimes SU(2)_L\otimes U(1)_Y$ \hspace{0.5cm} & \hspace{0.5cm} $U(1)_X$ \hspace{0.5cm} &\hspace{0.5cm} $U(1)_{xq-\tau_R^3}$ \hspace{0.5cm}&\hspace{0.5cm} $U(1)_{q+ \tilde{x} u}$ \hspace{0.5cm}\\
			\hline \hline
			$q_L^i$ \ \             & $(3, 2, \frac{1}{6})$\ \      & $x_q= \frac{1}{6}x_H + \frac{1}{3} x_{\Phi}$ & $x$ & $\frac{1}{3}$ \\[0.1cm]
            $u_R^i$ \ \             & $(3, 1,  \frac{2}{3})$\ \      & $x_u=\frac{2}{3}x_H + \frac{1}{3}x_{\Phi} $& $-1+4 x$ & $\frac{\tilde{x}}{3}$ \\[0.1cm]
            $d_R^i$ \ \             & $(3, 1, -\frac{1}{3})$\ \      & $x_d=-\frac{1}{3} x_H + \frac{1}{3} x_{\Phi} $ & $1-2x $ & $\frac{2-\tilde{x}}{3} $ \\[0.1cm]
            \hline \hline
			$\ell_L^i$ \ \             & $(1, 2, -\frac{1}{2})$\ \         &$x_\ell=-\frac{1}{2} x_H - x_{\Phi}$ &$-3x$ &$-1$ \\[0.1cm]
			$ e_R^i$ \ \        & $(1, 1, -1)$\ \                  & $x_e=-x_H - x_{\Phi}$ &$1-6x$ &$-(\frac{2+\tilde{x}}{3})$ \\[0.1cm]
            \hline \hline
			$H$       \ \  & $(1, 2, -\frac{1}{2})$\ \               & $-\frac{1}{2} x_H$ & $1-3x$ & $\frac{1-\tilde{x}}{3}$ \\[0.1cm]
			\hline \hline 
			$ N^j$ \ \           & $(1,1,0)$ \ \ 					    &$x_N=-x_{\Phi}$ &$-1$ &$\frac{-4+\tilde{x}}{3}$ \\[0.1cm]
			$\Phi$			\ \ &$(1,1,0)$ 				  \ \	    &$2\;x_{\Phi}$ &$2$ &$-2 (\frac{-4+\tilde{x}}{3})$ \\
			\hline \hline
		\end{tabular}
	\end{center}
	\caption{Particle content with the charge assignments of minimal $U(1)$ extensions of the SM. New SM-singlet RHNs $(N^j)$ and scalar $(\Phi)$ are included where $\{i, j\}$ represent three generations. Here $x_H$, $x_{\Phi}$, $x$ and $\tilde{x}$ are free real parameters.}
	\label{tab1}
\end{table} 
The Yukawa interactions consistent with the $\mathrm{SM}\otimes U(1)_X$ gauge symmetry are given by
\begin{align}
{\cal L}^{\rm Yukawa} ={}&
- Y_u^{\alpha \beta} \overline{q_L^\alpha} H u_R^\beta
- Y_d^{\alpha \beta} \overline{q_L^\alpha} \tilde{H} d_R^\beta
- Y_e^{\alpha \beta} \overline{\ell_L^\alpha}\tilde{H} e_R^\beta
\nonumber\\
&
- Y_\nu^{\alpha \beta} \overline{\ell_L^\alpha} H N_R^\beta
-\frac{1}{2}Y_N^\alpha \Phi \overline{(N_R^\alpha)^c} N_R^\alpha
+{\rm h.c.},
\label{LYk}
\end{align}
where $\tilde{H}=i\tau^2H^\ast$, with $\tau^2$ denoting the second Pauli matrix.
Following the spontaneous breaking of the $U(1)$ symmetry through the vacuum expectation value (VEV) of the singlet scalar $\Phi$, the new gauge boson $Z^\prime$ acquires mass, while the RHNs obtain Majorana masses through the last term of Eq.~(\ref{LYk}). Electroweak symmetry breaking subsequently generates the Dirac neutrino masses. Together, the Majorana and Dirac mass terms realize the type-I seesaw mechanism, thereby accounting for the observed tiny neutrino masses and flavor mixing.
\par Besides the general $U(1)_X$ framework, we also consider the
$U(1)_{xq-\tau_R^3}$ and $U(1)_{q+\tilde{x}u}$ realizations~\cite{Carena:2004xs,Hashimoto:2014ela}. Although these models possess the same chiral structure of the $Z^\prime$ interactions, they are distinguished by their fermion charge assignments, as summarized in Table~\ref{tab1}.

Fixing $x_\Phi=1$ in the $U(1)_X$ scenario without loss of generality, several well-known models are recovered for specific values of the parameter $x_H$. In particular, choosing $x_H=-2$ makes the $U(1)_X$ charges of the left-handed quark and lepton doublets vanish, reproducing the familiar $U(1)_R$ model in which only the right-handed SM fermions carry non-zero $U(1)_X$ charges. Similarly, the choices $x_H=-1$, $-1/2$, and $1$ render the charges of $e_R$, $u_R$, and $d_R$ zero, respectively, while the remaining fermions remain charged under $U(1)_X$. Since particles with vanishing $U(1)_X$ charges do not couple directly to the $Z^\prime$, these scenarios naturally exhibit chiral $Z^\prime$ interactions. In contrast, setting $x_H=0$ reproduces the well-known $U(1)_{B-L}$ model, where left and right-handed fermions carry identical $B-L$ charges and hence couple vectorially to the $Z^\prime$. In this limit, the SM Higgs doublet is neutral under $U(1)_{B-L}$, whereas the singlet scalar $\Phi$ carries a charge of $2$.
\par A similar pattern is observed in the $U(1)_{xq-\tau_R^3}$ realization. Choosing $x=0$ reproduces the $U(1)_R$ scenario, while the choices $x=1/6$, $1/4$, and $1/2$ make the charges of $e_R$, $u_R$, and $d_R$ vanish, respectively. The choice $x=1/3$ corresponds to the $U(1)_{B-L}$ limit.
For the $U(1)_{q+\tilde{x}u}$ realization, however, the left-handed lepton doublets remain charged for all values of $\tilde{x}$, preventing the realization of a pure $U(1)_R$ scenario. Instead, $\tilde{x}=0$, $2$, and $-2$ make the charges of $u_R$, $d_R$, and $e_R$ vanish, respectively, whereas $\tilde{x}=1$ again reproduces the $U(1)_{B-L}$ model.
\par The scalar sector is identical for all the $U(1)$ realizations considered here and is described by the renormalizable scalar potential
\begin{align}
V ={}&
m_h^2(H^\dag H)
+\lambda_H(H^\dag H)^2
+m_\Phi^2(\Phi^\dag\Phi)
+\lambda_\Phi(\Phi^\dag\Phi)^2
+\lambda^\prime(H^\dag H)(\Phi^\dag\Phi),
\label{pot1x}
\end{align}
where we assume the Higgs-portal coupling $\lambda^\prime$ to be sufficiently small such that the mixing between the doublet and singlet scalar mass eigenstates remains negligible.
After spontaneous symmetry breaking, the scalar fields develop the VEVs
\begin{align}
\langle H\rangle
&=
\frac{1}{\sqrt2}
\begin{pmatrix}
v+h\\
0
\end{pmatrix},
\qquad
\langle\Phi\rangle
=
\frac{v_\Phi+\phi}{\sqrt2},
\label{scalar-1}
\end{align}
where $v=246~{\rm GeV}$ is the electroweak VEV and $v_\Phi\gg v$ denotes the $U(1)$-breaking scale. Consequently, the mass of the new gauge boson is approximately given by $M_{Z^\prime}\simeq 2g_Xv_\Phi$,
where $g_X$ is the corresponding $U(1)$ gauge coupling. 
\subsection{Case-II}
An alternative realization of the general $U(1)$ extension is shown in
Table~\ref{tab2}. Unlike the minimal scenario discussed above, this model
contains an extended scalar sector consisting of two Higgs doublets and
three SM-singlet scalars together with three SM-singlet RHNs carrying non-universal $U(1)$ charges~\cite{Montero:2007cd}.
In particular, the first two RHNs, $N_{R_{1,2}}$, possess a different
$U(1)$ charge from $N_{R_3}$, while the SM fermions have universal
$U(1)$ charge assignments. The particle content and corresponding
charge assignments, satisfying all gauge anomaly cancellation conditions,
are summarized in Table~\ref{tab2}.
\begin{table}[h]
\begin{center}
\begin{tabular}{|c|c|c|c|c|c|}
\hline\hline
      &  $SU(3)_c$  & $SU(2)_L$ & $U(1)_Y$ & $U(1)_X$ \\ 
\hline
$q_{L_i}$ & {\bf 3 }    &  {\bf 2}         & $ 1/6$       &  $ x^\prime_q= (1/6) x_{H} + (1/3)$ \\
$u_{R_i}$ & {\bf 3 }    &  {\bf 1}         & $ 2/3$       & $x^\prime_u=(2/3) x_{H} + (1/3) $ \\
$d_{R_i}$ & {\bf 3 }    &  {\bf 1}         & $-1/3$       & $x^\prime_d=-(1/3) x_{H} + (1/3) $\\
\hline
\hline
$\ell_{L_i}$ & {\bf 1 }    &  {\bf 2}         & $-1/2$       & $x^\prime_\ell=(-1/2) x_{H} - 1 $ \\
$e_{R_i}$    & {\bf 1 }    &  {\bf 1}         & $-1$         & $x^\prime_e=-x_{H} - 1 $ \\
\hline
\hline
$N_{R_{1,2}}$    & {\bf 1 }    &  {\bf 1}         &$0$                    & $x^\prime_\nu=- 4 $ \\ 
$N_{R_3}$    & {\bf 1 }    &  {\bf 1}         &$0$                           & $x^{\prime\prime}_{\nu}=+ 5 $   \\
\hline
\hline
$H_1$            & {\bf 1 }    &  {\bf 2}         & $- 1/2$       & $(-1/2) x_{H}$ \\  
$H_2$            & {\bf 1 }       &  {\bf 2}       &$ -1/2$                  & $(-1/2) x_{H}+3 $  \\ 
$\Phi_1$            & {\bf 1 }       &  {\bf 1}       &$ 0$                  & $ +8  $  \\ 
$\Phi_2$            & {\bf 1 }       &  {\bf 1}       &$ 0$                  & $ -10 $  \\ 
$\Phi_3$          & {\bf 1 }       &  {\bf 1}       &$ 0$                  & $ -3 $  \\
\hline\hline
\end{tabular}
\end{center}
\caption{
Particle content of the `alternative' general $U(1)$ extension of the SM where $i$ denotes the generation index. 
}
\label{tab2}
\end{table}   
The Yukawa interactions allowed by the
$\mathrm{SM}\otimes U(1)_X$ gauge symmetry are
\begin{eqnarray}
- \mathcal{L}^{\text{lepton}}_Y &=& \bar{\ell}_L y_l \tilde{H}_1 e_R + \sum_{i=1}^3 \sum_{j=1}^2 Y_D^{ij} \bar{\ell}_{Li} H_2 N_{R_j} + \frac{1}{2}\sum_{k=1}^2 Y_{2}^{k} \bar{N}^C_{R_{k}}\Phi_1 N_{R_k} + \frac{1}{2} Y_3 \bar{N}^C_{R_3}\Phi_2 N_{R_3} + \text{H.c.} \nonumber \\
-\mathcal{L}^{\text{quark}}_Y &=& \bar{Q}_L y_d \tilde{H}_1d_R + \bar{Q}_L y_u H_1u_R  + \text{H.c.}
\label{LYukawa}
\end{eqnarray}
The roles of the scalar fields are clearly distinguished.
The Higgs doublet $H_1$ generates the masses of the charged leptons and
quarks, whereas the second Higgs doublet $H_2$ couples exclusively to
the lepton doublets and the first two generations of RHNs,
thereby generating their Dirac mass terms after electroweak symmetry
breaking. The singlet scalars $\Phi_1$ and $\Phi_2$ generate the
Majorana masses of $N_{R_{1,2}}$ and $N_{R_3}$, respectively, once the
$U(1)_X$ symmetry is spontaneously broken.
Because of the $U(1)_X$ charge assignment, a Dirac Yukawa coupling for
$N_{R_3}$ is forbidden. Consequently, $N_{R_3}$ does not participate in
the seesaw mechanism, and the light neutrino masses are generated solely
through the first two RHNs. Throughout this work we work in a basis in
which the Majorana Yukawa matrix $Y_2$ is diagonal. The scalar potential is given by
\bea
  V&\ =\ &
m_{H_1}^2 (H_1^\dagger H_1) + \lambda_{H_1}  (H_1^\dagger H_1)^2 + m_{H_2}^2 (H_2^\dagger H_2) + \lambda_{H_2}  (H_2^\dagger H_2)^2 \nonumber \\
&& + m_{\Phi_1}^2 (\Phi_1^\dagger \Phi_1) + \lambda_1  (\Phi_1^\dagger \Phi_1)^2 
+ m_{\Phi_2}^2 (\Phi_2^\dagger \Phi_2) + \lambda_2   (\Phi_2^\dagger \Phi_2)^2 \nonumber \\
&&+ m_{\Phi_3}^2 (\Phi_3^\dagger \Phi_3) + \lambda_3   (\Phi_3^\dagger \Phi_3)^2 
+ ( \mu \Phi_3 (H_1^\dagger H_2) + {\rm H.c.} )  \nonumber \\
&&+ \lambda_4 (H_1^\dagger H_1) (H_2^\dagger H_2)+ \lambda_5 (H_1^\dagger H_2) (H_2^\dagger H_1) +\lambda_6 (H_1^\dagger H_1) (\Phi_1^\dagger \Phi_1)\nonumber \\
&&+ \lambda_7 (H_1^\dagger H_1) (\Phi_2^\dagger \Phi_2)+ \lambda_8 (H_1^\dagger H_2) (\Phi_3^\dagger \Phi_3) +\lambda_9 (H_2^\dagger H_2) (\Phi_1^\dagger \Phi_1)  \nonumber \\
&&+ \lambda_{10} (H_1^\dagger H_1) (\Phi_2^\dagger \Phi_2)+ \lambda_{11} (H_1^\dagger H_2) (\Phi_3^\dagger \Phi_3)+  \lambda_{12} (\Phi_1^\dagger \Phi_1) (\Phi_2^\dagger \Phi_2) \nonumber \\
&&+ \lambda_{13} (\Phi_2^\dagger \Phi_2) (\Phi_3^\dagger \Phi_3)+ \lambda_{14} (\Phi_3^\dagger \Phi_3) (\Phi_1^\dagger \Phi_1).
\label{HiggsPotential-2}
\eea
The scalar fields develop the vacuum expectation values
\bea
  \langle H_1 \rangle \ = \  \frac{1}{\sqrt 2}\left(  \begin{array}{c}  
    v_{h_1} \\
    0 \end{array}
\right),   \; 
\langle H_2 \rangle \ = \   \frac{1}{\sqrt{2}} \left(  \begin{array}{c}  
    v_{h_2}\\
    0 \end{array}
\right),  
\langle \Phi_1 \rangle \ = \  \frac{v_{1}}{\sqrt{2}},  \; 
\langle \Phi_2 \rangle \ = \  \frac{v_{2}}{\sqrt{2}},  \; 
\langle \Phi_3 \rangle \ = \  \frac{v_{3}}{\sqrt{2}},~~~~ 
\eea   
subject to the electroweak constraint
$v_{h_1}^2+v_{h_2}^2=(246~{\rm GeV})^2$.
For simplicity, we assume the mixed quartic couplings between the Higgs doublets and the SM-singlet scalars to be negligibly small. Consequently, the singlet and doublet scalar sectors are approximately decoupled, suppressing higher-order mixing effects among the heavy neutrinos after the spontaneous breaking of the $U(1)_X$ symmetry. The two sectors communicate only through the trilinear interaction
$\mu \Phi_3(H_1^\dagger H_2)+{\rm H.c.}$.
Since collider constraints require
$v_1^2+v_2^2+v_3^2\gg v_{h_1}^2+v_{h_2}^2$,
this interaction has a negligible impact on determining the singlet VEVs. Throughout this work, we choose the scalar potential parameters such that
$v_1\simeq v_2\simeq v_3$
and $\mu<v_1$.
The singlet scalar $\Phi_3$ acts as a spurion for the Higgs-doublet sector, inducing the mixing between $H_1$ and $H_2$ through the trilinear term. After $\Phi_3$ acquires the VEV
$\langle\Phi_3\rangle=v_3/\sqrt{2}$,
the induced mixing parameter is $m_{\rm mix}^2=\mu v_3/\sqrt{2}$. The Higgs-doublet sector therefore effectively reduces to a neutrinophilic two-Higgs-doublet model (2HDM). Owing to the $U(1)_X$ symmetry, no bilinear mixing terms among the singlet scalars are allowed, leaving two physical Nambu--Goldstone (NG) bosons in the spectrum. In our analysis, we assume all singlet scalars to be heavier than the $Z^\prime$ boson so that the corresponding decay channels are kinematically forbidden.

After the spontaneous breaking of the $U(1)_X$ symmetry, the $Z^\prime$ mass is given by
\bea
 M_{Z^\prime}
 = g_X \sqrt{64 v_{1}^2+100 v_{2}^2+9v_3^2
 +\frac14 x_H^2v_{h_1}^2
 +\left(-\frac12x_H+3\right)^2v_{h_2}^2}
 \simeq
 g_X\sqrt{64v_1^2+100v_2^2+9v_3^2},
\label{masses-Alt}
\eea
where the approximate expression follows from
$v_{1,2,3}\gg v_{h_{1,2}}$.

The $U(1)_X$ charge assignments forbid Yukawa interactions between the Higgs doublet $H_1$ and the RHNs, while $H_2$ exclusively participates in neutrino mass generation. This realization therefore belongs to the class of neutrinophilic two-Higgs-doublet models (2HDMs)~\cite{Ma:2000cc,Wang:2006jy,Gabriel:2006ns,Davidson:2009ha,Haba:2010zi}. In the phenomenologically relevant limit $0<m_{\rm mix}^2=\mu v_3/\sqrt2\ll m_{\Phi_3}^2$,
the VEV of the neutrinophilic Higgs doublet satisfies $v_{h_2}\simeq
m_{\rm mix}^2 v_{h_1}/m_{\Phi_3}^2
\ll v_{h_1}$, naturally generating a small Dirac neutrino mass. The dependence on the parameter $x_H$ is identical to that of the minimal $U(1)_X$ model with $x_\Phi=1$. Throughout the remainder of this work, we refer to this framework as the {\it alternative} $U(1)$ scenario.
\subsection{Neutrino mass}
The spontaneous breaking of the general $U(1)$ symmetry in Case-I, summarized in Tab.~\ref{tab1}, generates the Majorana masses of the RHNs, $M_{N_\alpha}=Y_N^\alpha v_\Phi/\sqrt{2}$,
through Eq.~(\ref{LYk}). After electroweak symmetry breaking, the Dirac neutrino masses are generated as $m_D^{\alpha\beta}=Y_\nu^{\alpha\beta}v/\sqrt{2}$.
The resulting neutrino mass matrix takes the form
\begin{equation}
m_\nu=
\begin{pmatrix}
0 & m_D \\
m_D^T & M_N
\end{pmatrix},
\label{num-1-1}
\end{equation}
which, in the seesaw limit $(m_D\ll M_N)$, yields the light neutrino mass matrix
\begin{align}
m_\nu\simeq -m_D M_N^{-1}m_D^T.
\label{nut}
\end{align}
Consequently, the light and heavy neutrino mass eigenstates are mixed by an angle of order $m_D/M_N$, allowing the heavy neutrinos to interact with the SM gauge bosons through their small admixture with the active neutrinos. In this minimal realization, all three generations of RHNs participate in the seesaw mechanism and account for the observed neutrino masses and flavor mixing.

In the alternative scenario (Case-II), shown in Tab.~\ref{tab2}, the first two RHNs acquire Majorana masses $M_{N_{1,2}}=Y_2^{1,2} v_1/\sqrt{2}$, while the third generation obtains
$M_{N_3}=Y_3 v_2/\sqrt{2}$,
following the Yukawa interactions in Eq.~(\ref{LYukawa}). The Dirac mass matrix is generated solely through the neutrinophilic Higgs doublet $H_2$, $m_D^{ij}=Y_D^{ij} v_{h_2}/\sqrt{2}$, and involves only $N_{R_{1,2}}$. Consequently, only the first two RHNs participate in the type-I seesaw mechanism, whereas the $U(1)_X$ charge assignment forbids a Dirac Yukawa coupling for $N_{R_3}$. As a result, $N_{R_3}$ neither mixes with the active neutrinos nor contributes to neutrino mass generation at tree level, making it a viable dark matter candidate. The light neutrino masses are therefore generated through a minimal seesaw framework with two RHNs~\cite{Smirnov:1993af,King:1999mb,Frampton:2002qc,Ibarra:2003up}, described by Eqs.~(\ref{num-1-1}) and~(\ref{nut}).
\subsection{$Z^\prime$ interactions with the fermions}
Under general $U(1)$ scenario given in Case-I and II, the neutral BSM gauge boson $Z^\prime$ interacts with the SM fermions $(f_{L(R)})$ following
\bea
\mathcal{L} = -g_X (\overline{f_L}\gamma^\mu q_{f_{L}^{}}^{}  f_L+ \overline{f_R}\gamma^\mu q_{f_{R}^{}}^{}  f_R) Z_\mu^\prime.
\label{Lag1}
\eea
where $q_{f_{L(R)}^{}}^{}$ is general $U(1)$ charge of the left (right) handed fermions. We calculate the partial decay widths of $Z^\prime$ into corresponding charged leptons and quarks as
\begin{align}
\label{eq:width-ll}
\Gamma(Z^\prime \to \bar{f} f)
= N_C^{} \frac{M_{Z^\prime}^{} g_{X}^2}{24 \pi} \left[ \left( q_{f_L^{}}^2 + q_{f_R^{}}^2 \right) \left( 1 - \frac{m_f^2}{M_{Z^\prime}^2} \right) + 6 q_{f_L^{}}^{} q_{f_R^{}}^{} \frac{m_f^2}{M_{Z^\prime}^2} \right]
\sqrt{1-\frac{4 m_f^2}{M_{Z^\prime}^2}}~,
\end{align}    
where $m_f$ is the mass of the corresponding SM fermions and $N_C^{}=1(3)$ is the color factor for SM leptons(quarks). The partial decay width of $Z^\prime$ into a pair of single generation light neutrinos neglecting the tiny neutrino mass can be written as
\begin{align}   
\label{eq:width-nunu}
    \Gamma(Z^\prime \to \nu \nu)
    =  \frac{M_{Z^\prime}^{} g_{X}^2}{24 \pi} q_{f_L^{}}^2~,
\end{align} 
where $q_{f_L^{}}^{}$ is the general $U(1)$ charge of the SM lepton doublet $\ell_L$. In general $U(1)$ extended SM scenarios, $Z^\prime$ interacts with heavy neutrinos following 
\bea
\mathcal{L}_N= -\frac{1}{2}g_X q_{N_R} \overline{N} \gamma_\mu \gamma_5 N Z_{\mu}^\prime.
\label{neut}
\eea
where $q_{N_R}$ is the  general $U(1)$ charge of heavy neutrinos. The partial decay width of $Z^\prime$ into a pair of single generation heavy neutrino can be written as
\begin{align}
\label{eq:width-NN}
    \Gamma(Z^\prime \to N^\alpha N^\alpha)
    = \frac{ g_{X}^2 M_{Z^\prime}^{}}  {24 \pi} q_{N_R^{}}^2 \left( 1 - \frac{4 M_{N_\alpha}^2}{M_{Z^\prime}^2} \right)^{\frac{3}{2}}~,
\end{align}
where $M_{N_\alpha}$ being the heavy neutrino mass. These expressions of the partial decay widths of $Z^\prime$ are same for the general $U(1)$ scenarios given in Tab.~\ref{tab1} and \ref{tab2}. Depending on the model set up charges of the left and right handed fermions will differ. 
The total decay width of $Z^\prime$ is obtained as a function of $M_{Z^\prime}$, $x_H$ and $g_X$ and it can be used to calculate its decay length at rest frame as
\begin{align}
c\tau \,[\text{m}]= \frac{1.97\times 10^{-16}} {\Gamma[M_{Z^\prime}, g_X, x_H]\,[\text{GeV}]},
\end{align}
where $\Gamma[M_{Z^\prime}, g_X, x_H]$ is the total decay width of $Z^\prime$ inclusive of all partial decay widths. At FPF@FCC, long-lived $Z^\prime$ bosons are expected to be detectable if their decay length is approximately in the range of $1.5$--$2.0$ km, depending on the FPF1 or FPF2 experimental configuration.
\par The branching ratios of the $Z^\prime$ depend sensitively on the general $U(1)_X$ charge assignment through the chiral couplings to SM fermions and RHNs~\cite{A:2025ygb}. In the $U(1)_X$ scenario, the decay mode $Z^\prime\to NN$ can become dominant for certain $U(1)_X$ charge assignments, while for others it remains subdominant to visible leptonic and hadronic decay channels. In the alternative scenario, the larger $U(1)_X$ charges of the RHNs substantially enhance the $Z^\prime\to NN$ branching ratio, making it the dominant decay mode over a broad region of parameter space. Consequently, the production rate of RHNs and the sensitivity of forward LLP searches are strongly model dependent.
\subsection{Decay of heavy neutrinos}
SM-singlet RHNs (HNLs) do not interact with the SM sector directly. Due to the seesaw mechanism we can express the light neutrino $(\nu)$ flavor eigenstate in terms of the light $(\nu_m)$ and heavy $(N_m)$ neutrino mass eigenstates as 
\bea 
  \nu \simeq  \nu_m  + V_{\ell N} N_m~,  
\eea 
where mixing between the light and heavy mass eigenstates can be written as $V_{\ell N}=m_D/m_N$ assuming $|V_{\ell N}| \ll 1$. Implementing flavor eigenstates in terms of the mass eigenstates, we write the charged current (CC) interaction as
\bea 
\mathcal{L}_{\rm CC} \supset 
 -\frac{g}{\sqrt{2}} W_{\mu}
  \bar{\ell} \gamma^{\mu} P_L   V_{\ell N} N_m  + {\rm H.c.}, 
\label{CC}
\eea
where $\ell$ denotes the three generations of the charged leptons from the SM in the vector form, and 
$P_L =\frac{1}{2} (1- \gamma_5)$ stands for the projection operator. The neutral current (NC) interaction can be written as 
\bea 
\mathcal{L}_{\rm NC} \supset 
 -\frac{g}{2 \cos\theta_{\rm W}}  Z_{\mu} 
\left[ 
  \overline{N_m} \gamma^{\mu} P_L  |V_{\ell N}|^2 N_m 
+ \left\{ 
  \overline{\nu_m} \gamma^{\mu} P_L V_{\ell N}  N_m 
  + {\rm H.c.} \right\} 
\right] , 
\label{NC}
\eea
where $\theta_{\rm W}$ is the Weinberg mixing angle. Through the CC interaction in Eqs.~\eqref{CC} and the NC interaction in Eq.~\eqref{NC}, the heavy neutrino decay into the SM particles via light-heavy neutrino mixing. Depending on the mass of the heavy neutrinos, they can decay into two-body $(M_N > M_W, M_Z)$ and three-body $(M_N < M_W, M_Z)$ modes where $M_{W(Z)}$ is the $W(Z)$ boson mass of the SM. In this paper we consider the case where heavy neutrinos are much lighter than the SM gauge bosons manifesting various three-body decay modes via the virtual decays of the $W$ and $Z$ bosons. Heavy neutrinos decay into purely leptonic modes such as $N \to \nu \ell^+ \ell^-$ and $N \to 3\nu$ with charged $(\ell^{\pm})$ and neutral $(\nu)$ leptons where virtual SM gauge bosons decay leptonically. If the SM gauge bosons decay hadronically then semileptonic modes such as $N \to \nu \mathscr{H}^{0}$ and $N \to \ell^\pm \mathscr{H}^{\mp}$ with neutral (charged) ($\mathscr{H}^{0(\pm)}$) meson are manifested. The details of RHN decay channels can be found in Appendix.~\ref{appl:decay_width}. The total decay width of heavy neutrino is obtained as a function of $M_{N}$ and $|V_{\ell N}|^2$ and it can be used to calculate its decay length at rest frame as
\bea
c\tau\, [\text{m}]= \frac{1.97\times 10^{-16} }{\Gamma[M_N, |V_{\ell N}|^2]\, [\text{GeV}]},
\eea
where $\Gamma[M_N, |V_{\ell N}|^2]$ is the total decay width of heavy neutrinos inclusive of all partial decay widths. In the mass range of interest, the HNL is typically long-lived. FPF@FCC is sensitive to long-lived HNL with decay lengths of roughly $1.5$--$2.0$ km, depending on the FPF1(2) experimental setup.
\par The RHN branching ratios are primarily determined by their mass~\cite{A:2025ygb}. For RHN masses below the pion threshold, the invisible three-neutrino mode dominates. Once hadronic channels become kinematically accessible, semileptonic decays rapidly take over, leading to visible branching fractions exceeding $90\%$ for RHNs heavier than light mesons. Consequently, long-lived RHNs predominantly decay into visible final states over most of the relevant parameter space, making them promising targets for LLP searches at forward experiments.
\section{Long-lived particle searches and limits on model parameters}
\label{sec:experiment}
\begin{figure}[htb!]
\centering
\includegraphics[scale=0.36]{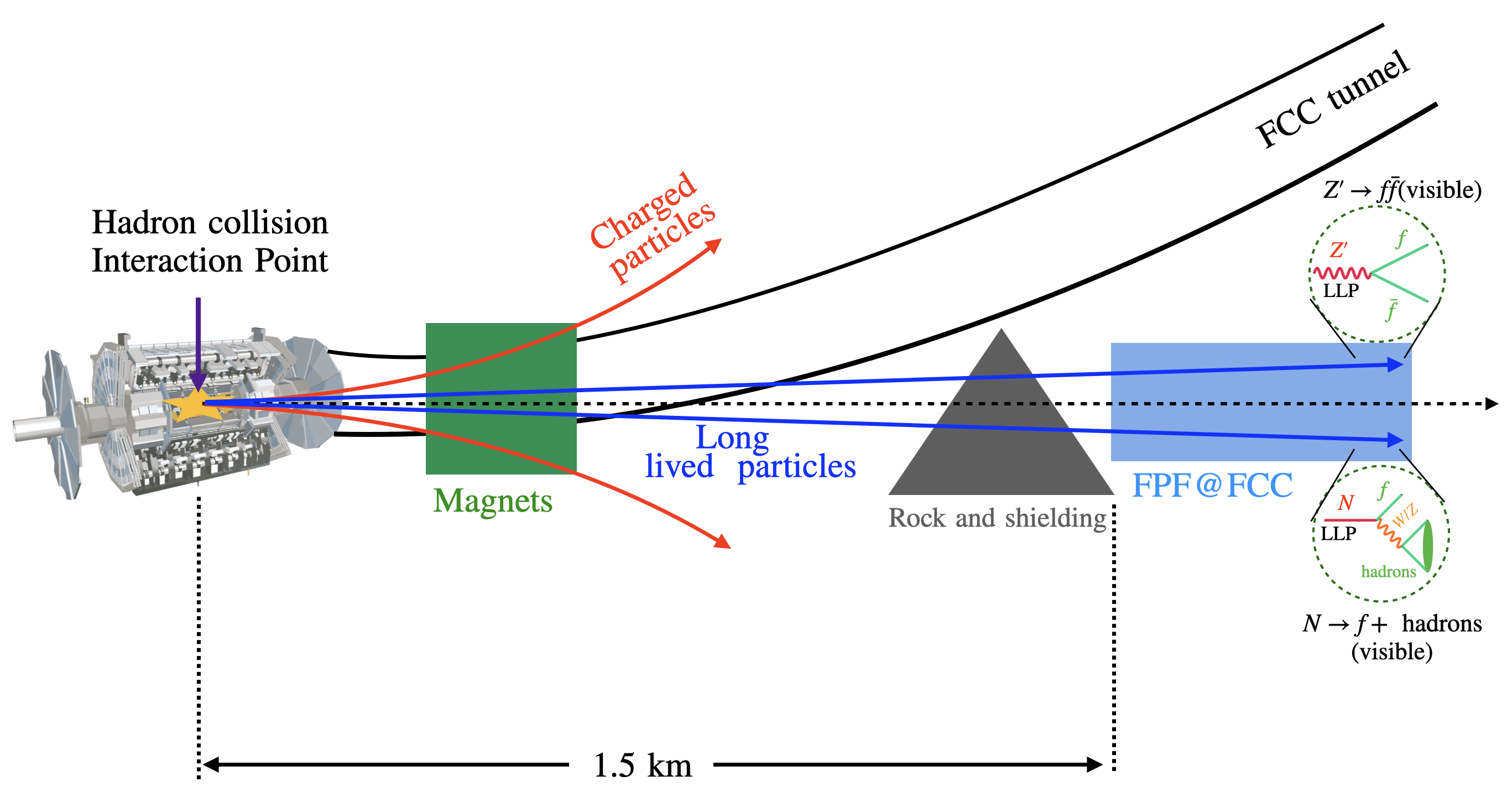}
\caption{Schematic representation of the production of LLPs at the FCC-hh IP at $\sqrt{s}=$100 TeV. The FPF@FCC detector set-up is 1.5 km away from the IP of the $pp$ collisions. We consider four cases here: (i) long-lived heavy neutrino production from meson decay, (ii) long-lived $Z^\prime$ production from meson decay and bremsstrahlung following its decay into visible modes when heavy neutrinos are kinematically inaccessible, (iii) $Z^\prime$ is long-lived and decay into visible modes inside the decay volume. In this case heavy neutrinos are long-lived and decay outside the detector and (iv) $Z^\prime$ is short lived and decay into long lived heavy neutrino pair followed by the visible decay of the heavy neutrinos inside the decay volume.}
\label{fig:llp1}
\end{figure}

\begin{figure}[htb!]
\centering
\includegraphics[scale=0.3]{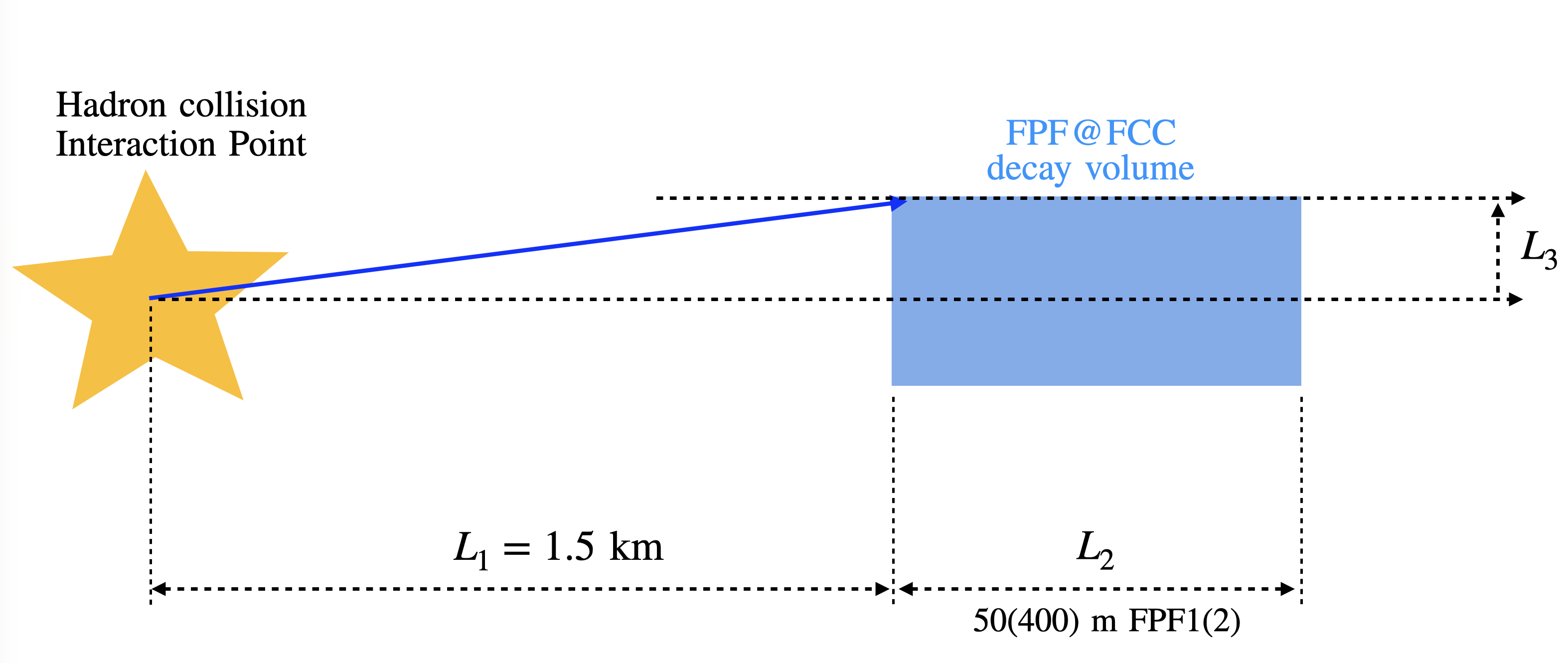}
\caption{Schematic representation of the experiment and FPF1(2) detector. The detector is $1.5$ km $(L_1)$ away from the FCC-hh IP whereas the detector length is 50(400) m $(L_2)$ and $L_3=2.5 (10)$ m.}
\label{fig:llp2}
\end{figure}

To explore the sensitivity of the FPF@FCC to LLPs, we consider two benchmark of FPF configurations, FCC-LLP1~(FPF1) and FCC-LLP2~(FPF2):
\bea
\text{\textbf{FPF1(2)}}:L_1 =1.5~ \text{km},~L_2 = 50 (400)~\text{m},~2L_3 = 5 (20)~\text{m},~  \mathcal{L} = \text{30~ab}^{-1}.
\eea
A schematic diagram of the experimental set-up has been shown in Fig.~\ref{fig:llp1} and \ref{fig:llp2}.
Both detector configurations are centered on the beam line of sight, with their upstream faces located 1.5 km downstream of the interaction point (IP). FPF1, adopted as the baseline setup, consists of a $5~\mathrm{m}\times5~\mathrm{m}\times50~\mathrm{m}$ decay volume, while FPF2 represents an extended configuration with dimensions of $20~\mathrm{m}\times20~\mathrm{m}\times400~\mathrm{m}$. Both detectors are assumed to operate at the FCC-hh with an integrated luminosity of $30~\mathrm{ab}^{-1}$ and a visible-energy threshold of $E_{\rm vis}\gtrsim100$ GeV. Their location, approximately 500m of rock shielding from the IP, together with charged-particle vetoes and precision tracking, provides a low-background environment for reconstructing displaced vertices from LLP decays. An electromagnetic calorimeter placed downstream of the decay volume further enhances the identification of electrons, positrons, and photons, extending the sensitivity to a broad class of visible LLP signatures.
\par As illustrated in Fig.~\ref{fig:llp2}, the detector is positioned 1.5 km downstream of IP, resulting in a narrow angular acceptance. Since LLPs are typically produced with finite angles $\theta$ relative to the beam axis, only those emitted within the detector acceptance can reach the detector, and an even smaller fraction decay within the fiducial volume. If an LLP of mass of $m_{\rm LLP}$ is produced at the FCC-hh IP with a momentum $p= |\vec{p}_{\rm lab}|$, then its probability of decay inside the FPF1(2) detector volume can be written in a general form as
\bea
    \mathcal{P}(p, \theta)&=&(e^{-L_1/d_{\rm lab}} - e^{-(L_1 + L_2)/d_{\rm lab}})\Theta(X/2- x(L_1))\Theta(Y/2- y(L_1)) ,\nonumber \\
    &\approx& \frac{L_2}{d_{\rm lab}}e^{-L_1/d_{\rm lab}}\Theta(X/2- x(L_1))\Theta(Y/2- y(L_1)),
\eea
where $X$ and $Y$ are the lengths of the cross-sectional area of the detector which in our case is $X=Y= 2 L_3$ as shown in Fig.~\ref{fig:llp2}. $\Theta$ stands for the Heaviside step function.
Here $d_\text{lab}$ denotes the decay length of the LLP in the lab frame, and it can be expressed as $d_\text{lab} = c\tau (p/m_{\rm LLP})$, with $\tau(m_{\rm LLP})$ is the lifetime(mass) of the LLP. We determine the planar co-ordinates for the LLP as $(x,y)$ after leaving FCC-hh IP with a velocity $\vec{v}_\text{lab}$ in the laboratory frame with a polar angle $\theta= \tan^{-1}\Big( \frac{{p_{\rm lab}}^\perp}{{p_{\rm lab}}_z}\Big)$ with ${p_{\rm lab}}^\perp=\sqrt{{p_{\rm lab}}_x^2+{p_{\rm lab}}_y^2}$ and $t=D/(|\vec{v}_{\rm lab}| \cos\theta)$ is the time required for the LLP to reach at the detector. Therefore we determine the co-ordinates as 
\bea
x(D) = \frac{D}{|\vec{v}_\text{lab}|\cos\theta}|\vec{v}_\text{lab}|\sin\theta \cos\phi, \,\, \, \,\,\,\, y(D) = \frac{D}{|\vec{v}_\text{lab}|\cos\theta} |\vec{v}_\text{lab}|\sin\theta \sin\phi,
\eea
taking the mass of the LLP we convert the co-ordinates in terms of $\vec{p}_{\rm lab}$ as 
\bea
x(D) = \frac{D}{|\vec{p}_\text{lab}|\cos\theta}|\vec{p}_\text{lab}|\sin\theta \cos\phi, \,\, \, \,\,\,\, y(D) = \frac{D}{|\vec{p}_\text{lab}|\cos\theta} |\vec{p}_\text{lab}|\sin\theta \sin\phi,
\eea
where $D$ is the distance between the FCC-hh IP and the detector and in our case $D= L_1$ according to Fig.~\ref{fig:llp2}. Here $\phi$ is the azimuthal angle determines as $\tan\phi=\frac{{p_{\rm lab}}_y}{{p_{\rm lab}}_x}$ where ${p_{\rm lab}}_{y(x)}$ is the $y(x)-$ laboratory frame component of the momenta of the LLPs. Now the fraction of produced LLPs decaying inside the detector volume FPF@FCC is evaluated by the acceptance Acc(LLP, $p,\, \theta$) and it can be expressed as
\begin{equation}
    \text{Acc}(\text{LLP}, p, \theta) = \mathcal{P}(p,\theta)~\text{BR}({\,\text{LLP}}\to \text{visible})~,
    \label{fig:angcut}
\end{equation}
Hence the number of signal events using the Acceptance factor is then formulated as 
\begin{align}
    N_S=\mathcal{L}\int dp d\theta \frac{d\sigma_{pp\to\text{LLP}+X}}{dpd\theta} \times  \text{Acc}(\text{LLP}, p, \theta).
\label{eq:Ns}
\end{align}
To generate the events there is a trigger requirement on the LLPs such that $p\gtrsim 100$~GeV~(or equivalently $E_{\rm vis}>100$ GeV) which is required for the LLPs traveling close to the beam collision axis and LLPs are sufficiently boosted to decay in FPF1(2) detector. At the FCC-hh IP different types of mesons  will be produced in high amount by $pp$ collisions with a characteristic momentum transfer scale at $p_T=\Lambda_{\rm QCD}\simeq$ 0.25 GeV. We expect high multiplicity of mesons with momentum  $p>100$ GeV will be produced at small angles corresponding to $\tan\theta \simeq L_3/L_1=2.5 (10)/1500$ for FPF1(2) detector. 
\par We estimate the projected sensitivity of FPF@FCC, considering the FCC-LLP1(2) detector configurations, which are shielded by rock and concrete. Cosmic-ray backgrounds are expected to be effectively suppressed by exploiting the directionality and timing of the reconstructed LLP signals. This allows us to adopt a conservative background-free assumption. In addition, we assume a $100\%$ detection efficiency for all visible decay modes.
\par We first consider long-lived heavy neutrinos produced from meson decays and, using their visible final states, derive limits in the $|V_{\ell N}|^2$--$M_N$ plane for $\ell=e,\mu$. We then study long-lived $Z^\prime$ bosons decaying visibly inside FCC-LLP1(2), assuming the heavy-neutrino channel is kinematically forbidden, $M_N>M_{Z^\prime}/2$. We also analyze the case where a long-lived $Z^\prime$ decays into both visible SM final states and long-lived heavy neutrinos with $M_N=M_{Z^\prime}/3$, for which the heavy neutrinos typically decay outside the detector. Finally, we consider promptly decaying $Z^\prime$ bosons producing long-lived heavy neutrinos, which subsequently decay visibly inside FCC-LLP1(2). From the $Z^\prime$-mediated channels, we derive bounds in the $g_X$--$M_{Z^\prime}$ plane for different general $U(1)$ charge assignments, while the prompt-$Z^\prime$ scenario yields limits in the $|V_{\ell N}|^2$--$M_N$ plane. In this analysis we used the DarkCast \cite{Baruch:2022esd} package to calculate branching ratios of $Z^\prime$ in the chiral scenarios using Eqs.~(\ref{eq:width-ll}), (\ref{eq:width-nunu}), (\ref{eq:width-NN}) under consideration and FORESEE \cite{Kling:2021fwx} package for event analyses for FPF@FCC. Partial decay widths of heavy neutrinos have been used from Appendix \ref{appl:decay_width}. We discuss our analyses and adopted scenarios in the following:
\subsection{Long-lived heavy neutrinos from meson decays}
\begin{figure*}[htb!]
\centering
\includegraphics[scale=0.375]{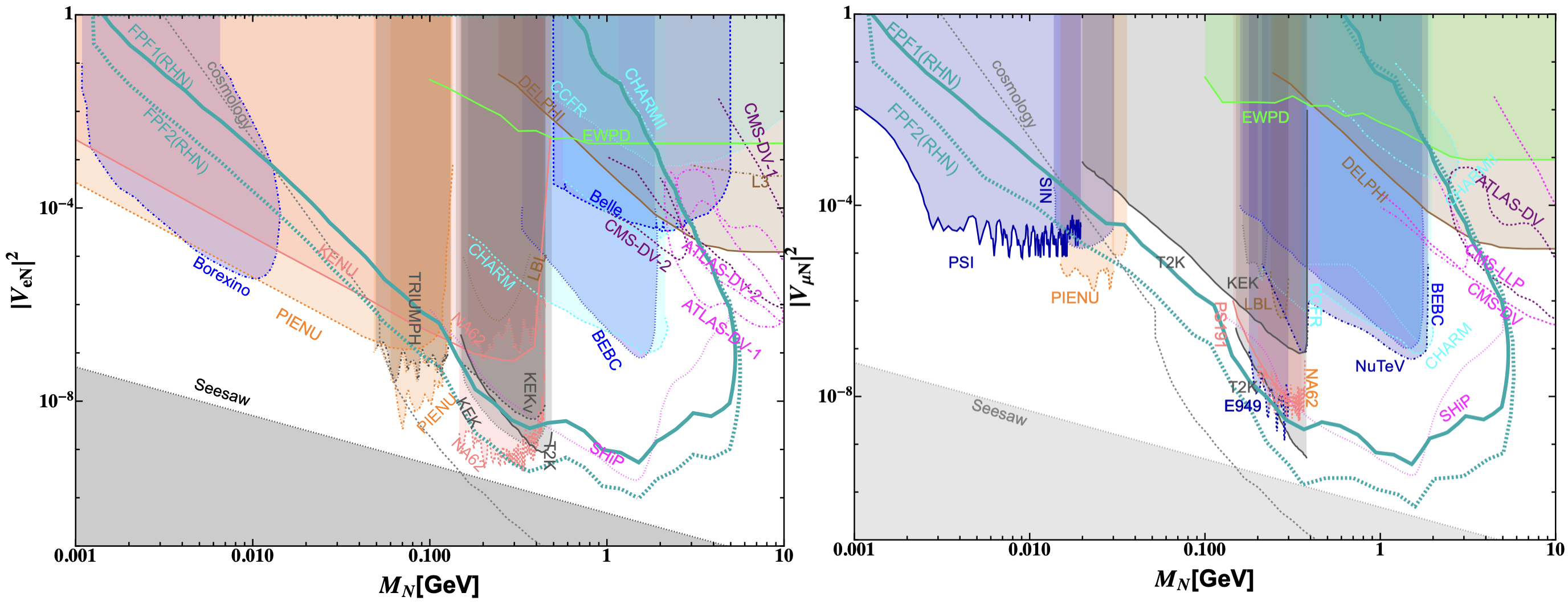}
\caption{Prospective limits on the light--heavy neutrino mixing as a function of $M_N$ for long-lived heavy neutrinos produced in meson decays. The solid (dashed) dark-cyan curve corresponds to the FPF1(RHN) [FPF2(RHN)] configuration, considering the visible decay channels $N\to \ell/\nu_\ell+\mathrm{X}$. The shaded regions are excluded by existing experimental constraints and theoretical estimations.}
\label{fig:mixing}
\end{figure*}

We are interested in HNLs produced in meson decays which are triggered by active-heavy neutrino mixing $V_{\ell N}$. Given their large production rates at the FCC-hh, we focus on pseudoscalar mesons including $B^\pm$, $B_c^\pm$, $B^0$, $D^\pm$, $D_s^\pm$, $D^0$, $K^\pm$, $\pi^\pm$. We considered HNL production both from the two-body and three-body decays of these mesons. Owing to their long-lived nature, these heavy neutrinos can reach the FPF1(2) detector and decay into visible final states within the detector acceptance.
\begin{equation}
    \text{Acc}(\text{LLP}, p_N, \theta_N) = \mathcal{P}(p_N,\theta_N)~\text{BR}(N \to \text{visible})~,
    \label{fig:angcut}
\end{equation}
The differential production rate of heavy neutrino from meson decay can be given by
\begin{align}
\label{eq:prod-rate_p-meson}  
    \frac{\dd N^M}{\dd p_M^2 \dd \cos\theta_M^{}} =
    \frac{d\sigma(p p \to M X)}{\dd p_M^2 \dd \cos\theta_M^{}} \cdot 
    {\rm BR}(M \to N Y),
\end{align}
where ${\rm BR}(M \to N Y)$  includes only the two-body meson decay modes. In the present analysis, we additionally include the corresponding three-body decay channels. Consequently, the expected number of heavy-neutrino decay events in the FPF1(2) detector, obtained using Eqs.~(\ref{fig:angcut}) and (\ref{eq:prod-rate_p-meson}), is
\begin{align}
    \label{eq:num_p-meson}
    N_{\rm event}^{\rm p-meson}& =
    N_p\, \sum_{M} \int \dd p_M^2 \int \dd \cos\theta_M^{} \int \dd p_N^2 \int \dd \cos\theta_N^{}\, \frac{\dd N^M}{\dd p_M^2 \dd \cos\theta_M^{}} \cdot {\rm Acc}(\text{LLP},p_N,\theta_N).
\end{align}
Assuming a $95\%$ confidence-level sensitivity corresponding to
$N_{\rm event}^{\rm p\text{-}meson}>3$ in
Eq.~(\ref{eq:num_p-meson}), we derive projected bounds on the
light--heavy neutrino mixing $|V_{e(\mu)N}|^2$ as a function of $M_N$.
The resulting sensitivities are shown in the left (right) panel of
Fig.~\ref{fig:mixing} for electron- (muon-) flavored HNLs, using visible
decay channels of the form
$N\to e(\mu)/\nu_{e(\mu)}+\mathrm{X}$ mediated by charged- and
neutral-current interactions in the FPF1 and FPF2 detectors. These lines are represented by  solid dark cyan (FPF1(RHN)) and dashed (FPF2(RHN)) lines.
\par As shown in Fig.~\ref{fig:mixing}, analyzing the visible final state from the process $N\to \ell/ \nu_{\ell}+$ associated particles, we find that FPF1 may provide a strong limit on the light-heavy neutrino mixing around $|V_{\ell N}|^2 \leq 10^{-9} - 10^{-5}$ for the heavy neutrino mass range $M_N\sim 0.5-6$ GeV, whereas for lighter mass range 0.1 GeV $\leq M_N \leq 0.3$ GeV, the sensitivity lies in the range $|V_{\ell N}|^2 \leq 10^{-8} - 10^{-6}$. For the same mass range, these limits could be improved by $\mathcal{O}(1)$ magnitude in FPF2 
as shown by dashed cyan line. Similar to the SHiP experiment~\cite{Alekhin:2015byh,SHiP:2018xqw}, the strongest constraints originate from heavy neutrinos produced in pion, kaon, charm-meson, and $B$-meson decays, corresponding approximately to the mass ranges $M_N\lesssim0.14~{\rm GeV}$, $0.14~{\rm GeV}\lesssim M_N\lesssim0.45~{\rm GeV}$, $0.5~{\rm GeV}\lesssim M_N\lesssim2.5~{\rm GeV}$, and $2.5\text{ GeV }\lesssim M_N\lesssim 6~{\rm GeV}$, respectively. Although the FPF1 setup closely resembles SHiP, it offers improved sensitivity for $M_N \gtrsim 2.5$ GeV owing to the substantially larger production rate of $B$ mesons at the FPF. Furthermore, the larger detector volume of FPF2 enables it to surpass the projected SHiP sensitivity over the entire mass range considered.
\par In the left panel of Fig.~\ref{fig:mixing}, we compare our projected sensitivity with the existing experimental constraints on $|V_{eN}|^2$. For sub-kaon masses, peak-search experiments at TRIUMPH \cite{Britton:1992xv} and PIENU \cite{PIENU:2017wbj,Bryman:2019bjg,PIENU:2019usb} constrain the light-heavy neutrino mixing through the decay $\pi^+\to e^+N$ in the mass range $10^{-3}~{\rm GeV}\leq M_N\leq0.15~{\rm GeV}$. Within $0.003~{\rm GeV}\leq M_N\leq0.016~{\rm GeV}$, the Borexino experiment \cite{Borexino:2013bot} provides the most stringent limits by searching for the decay $N\to\nu e^+e^-$. For masses above the pion threshold, the NA62 peak-search analysis \cite{NA62:2020mcv,NA62:2025csa} using $K^+\to e^+N$ decays sets the strongest constraints in the range $0.15~{\rm GeV}\leq M_N\leq0.45~{\rm GeV}$. In the intermediate mass region, the T2K near detector \cite{T2K:2019jwa} probes heavy neutrinos produced via charged-current interactions and decaying in flight, providing competitive limits for $0.4~{\rm GeV}\leq M_N\leq0.5~{\rm GeV}$.
For heavier neutrino masses, beam-dump experiments provide the leading constraints. The BEBC \cite{WA66:1985mfx,Barouki:2022bkt} and CHARM \cite{CHARM:1985nku} experiments exclude sizeable light-heavy mixing in the ranges $0.5~{\rm GeV}\leq M_N\leq2~{\rm GeV}$ and $2~{\rm GeV}\leq M_N\leq2.25~{\rm GeV}$, respectively. At higher masses, searches at Belle \cite{Zhou:2021ylt}, DELPHI \cite{DELPHI:1996qcc}, ATLAS \cite{Tastet:2021vwp,ATLAS:2022atq}, and CMS \cite{CMS:2022fut} provide the most stringent exclusion limits over the range $2.25~{\rm GeV}\leq M_N\leq16~{\rm GeV}$. We further compare our projections with the latest displaced-vertex searches from NA62 \cite{NA62:2020mcv}, ATLAS \cite{ATLAS:2019kpx,ATLAS:2022atq,ATLAS:2024fdw}, and CMS \cite{CMS:2024xdq,CMS:2024bni}, as well as the projected sensitivity of the proposed SHiP experiment~\cite{Alekhin:2015byh,SHiP:2018xqw}.
\par In the right panel of Fig.~\ref{fig:mixing}, we compare our projected sensitivity with the existing constraints on $|V_{\mu N}|^2$. For sub-pion masses, heavy neutrinos produced in pion decays are constrained by the PSI \cite{Daum:1987bg} and PIENU \cite{PIENU:2019usb} experiments in the ranges $1.2\times10^{-3}~{\rm GeV}\leq M_N\leq0.0175~{\rm GeV}$ and $0.0175~{\rm GeV}\leq M_N\leq0.025~{\rm GeV}$, respectively. The T2K experiment \cite{Arguelles:2021dqn} also provides competitive limits through searches for heavy neutrino decays in the near detector. For heavier neutrinos, MicroBooNe \cite{Kelly:2021xbv} and KEK \cite{Hayano:1982wu} constrain the mass range $0.0325~{\rm GeV}\leq M_N\leq0.375~{\rm GeV}$, while BNL \cite{BNL-E949:2009dza}, BEBC \cite{WA66:1985mfx}, NuTeV \cite{NuTeV:1999kej}, CHARM \cite{CHARM:1985nku}, NA62 \cite{NA62:2025csa}, ATLAS \cite{ATLAS:2019kpx,ATLAS:2022atq}, and CMS \cite{CMS:2022fut} provide the most stringent constraints over the range $0.2~{\rm GeV}\leq M_N\leq11~{\rm GeV}$. We also compare our projections with the projected sensitivity of the proposed SHiP experiment \cite{SHiP:2018xqw}.
Combined cosmological constraints from CMB and BBN are taken from Refs.~\cite{Vincent:2014rja,Dolgov:2000pj,Ruchayskiy:2012si,Gelmini:2020ekg,Langhoff:2022bij,Sabti:2020yrt,Boyarsky:2020dzc}, providing indirect bounds on $|V_{e(\mu)N}|^2$ for $M_N\lesssim0.2~{\rm GeV}$, shown by the gray dot-dashed curve in Fig.~\ref{fig:mixing}. For comparison, the gray dashed line labeled ``Seesaw'' denotes the theoretical lower bound obtained from the seesaw relation assuming $m_\nu\leq0.1$ eV.
\par In summary, a comparison with existing experimental constraints and projected sensitivities shows that FPF1 can probe light-heavy neutrino mixing with a sensitivity comparable to the projected SHiP reach, while outperforming SHiP for $2.5~{\rm GeV}\lesssim M_N\lesssim6~{\rm GeV}$ owing to the enhanced production of $B$ mesons at the FPF. Benefiting from its larger detector volume, FPF2 further extends the accessible parameter space, surpassing both the current experimental bounds and the projected sensitivities of future experiments for $0.5~{\rm GeV}\lesssim M_N\lesssim 6~{\rm GeV}$~($0.04~{\rm GeV}\lesssim M_N\lesssim 6~{\rm GeV}$) in the $|V_{eN}|^2$~($|V_{\mu N}|^2$) plane.
\subsection{$Z^\prime$ induced scenarios}
Next, we consider three additional scenarios involving both long-lived and
short-lived $Z^\prime$ bosons. Depending on the available decay channels,
the $Z^\prime$ may decay into visible final states (leptons or hadrons) or,
if kinematically allowed, into a pair of heavy neutrinos. The $Z^\prime$
boson can be produced through the decays of $\pi^0$ and $\rho$ mesons,
with its mass constrained by the mass of the parent meson. The differential
production rate of $Z^\prime$ from meson decays is given by
\begin{align}
\label{eq:prod-rate_p-meson}  
    \frac{\dd N^M}{\dd p_M^2 \dd \cos\theta_M^{}} =
    \frac{d\sigma(p p \to M X)}{\dd p_M^2 \dd \cos\theta_M^{}} \cdot {\rm BR}(M \to Z' \gamma)~,
\end{align}
where $p_M^{}$ and $\theta_M^{}$ are the momentum and angle of meson respect to the beam axis. Additionally $Z^\prime$ can also be produced from the proton bremsstrahlung unrestricted by the meson masses. We calculate the $Z^\prime$ production \cite{Kim:1973he,Feng:2017uoz,Bauer:2018onh,Asai:2022zxw} from the bremsstrahlung process in the following way
\begin{align}
\label{eq:xsec-pbrems}
    \sigma(p p \to p Z' X) =
    \int \dd p_{Z'}^2 \int \dd \cos\theta_{Z'}^{} \frac{p_{Z'}}{p_{p_i}^{}} w(p_{Z'}^2, \cos\theta_{Z'}^{}) \sigma_{pp}(s')~, 
\end{align}
where $\theta_{Z'}$ stands for the angle of $Z'$ respect to the beam axis, $p_{p_i}^{}$ is the momentum of initial proton in the beam, $s' = 2 m_p (E_p - E_{Z'}^{})$ where $m_p$ represents the proton mass and $E_p~(E_{Z'}^{})$ is the energy of the initial proton ($Z'$). In Eq.~(\ref{eq:xsec-pbrems}) the inelastic $pp$ scattering cross-section is represented by $\sigma_{pp}(s)$ whereas $w(p_{Z'}^2, \cos\theta_{Z'}^{})$ represents the splitting function \cite{Kim:1973he,Tsai:1973py,Blumlein:2013cua,Asai:2022zxw}. The differential production rate of the $Z'$ gauge boson can be estimated as
\begin{align}
\label{eq:prod-rate_p-brem}
    \frac{\dd N^{\rm brem}}{\dd p_{Z'}^2 \dd \cos\theta_{Z'}^{}} =
    \frac{p_{Z'}^{}}{p_{p_i}} w(p_{Z'}^2, \cos\theta_{Z'}^{}) \frac{\sigma_{pp}(s')}{\sigma_{pp}(s)}~,
\end{align}
where $s = 2 m_p E_{p_i}$.
After production, these long-lived $Z^\prime$ gauge bosons travel to the FPF1(2) detector passing through shielding like rock and concrete. We consider the visible decay of $Z^\prime$ in our analysis according to  Fig.~\ref{fig:llp1} in the following way:
\subsubsection{Long-lived $Z^\prime$ and inaccessible heavy neutrinos}
\begin{figure*}[htb!]
\centering
\includegraphics[scale=0.43]{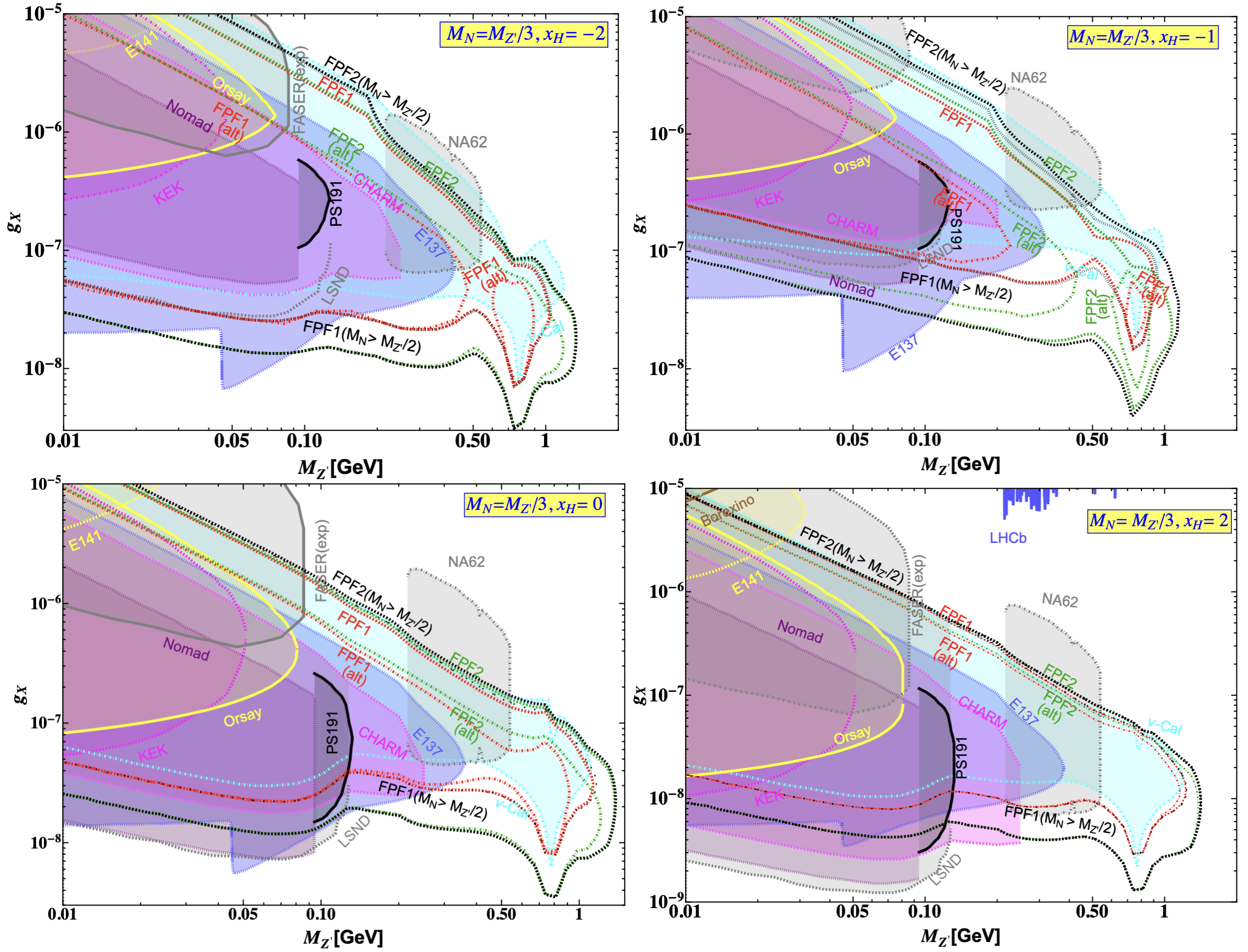}
\caption{Prospective limits on the general $U(1)$ gauge coupling as a function of
$M_{Z^\prime}$ for a long-lived $Z^\prime$ boson. The black dotted (dashed)
curve corresponds to the scenario in which heavy-neutrino decays are
kinematically forbidden, $M_N > M_{Z^\prime}/2$, for FPF1 (FPF2). The red
(dark-green) dashed curves show the sensitivities for kinematically
accessible long-lived heavy neutrinos with
$M_{N_{1,2}} = M_{Z^\prime}/3$ in the general $U(1)$ scenario, while the
corresponding alternative (\textit{alt}) scenarios are represented by the
red (dark-green) dot-dashed curves for FPF1 (FPF2). In the latter cases,
the mixing is fixed to $|V_{\ell N}|^2 = 10^{-6}$. The shaded regions are
excluded by existing experimental constraints.}
\label{params3}
\end{figure*}
We consider the case where $Z^\prime$ is long-lived
and decays into visible modes consisting of hadrons and charged leptons inside FPF1(2). We assume $M_N > M_{Z^\prime}/2$ so that heavy neutrinos are kinematically inaccessible. Therefore visible signal in this scenario consists of the visible decay modes of $Z^\prime$ only. Hence the corresponding acceptance can be evaluated as
\begin{equation}
\text{Acc}(\text{LLP}, p_{Z'}, \theta_{Z'}) = \mathcal{P}(p_{Z'},\theta_{Z'})~\text{BR}(Z'\to \text{visible})~.
\label{eq:accep_0}
\end{equation}
From Eqs.~\eqref{eq:prod-rate_p-meson}, (\ref{eq:prod-rate_p-brem}) and (\ref{eq:accep_0}), the total expected numbers of events in FPF1(2) from meson decay and bremsstrahlung is given by
\begin{align}
\label{eq:num_p-meson1}
N_{\rm event}^{\rm p\mathchar`-meson}& =
    N_p\, \sum_{M = \pi^0, \eta} \int \dd p_M^2 \int \dd \cos\theta_M^{} \int \dd p_{Z'}^2 \int \dd \cos\theta_{Z'}^{}\, \frac{\dd N^M}{\dd p_M^2 \dd \cos\theta_M^{}} \cdot {\rm Acc}(\text{LLP},p_{Z'},\theta_{Z'})~,\\
\label{eq:num_p-brem1}
N_{\rm event}^{\rm p\mathchar`-brem}& =
    N_p\, |F_1(m_{Z'}^2)|^2 \int \dd\, p_{Z'}^2 \int \dd \cos\theta_{Z'}^{}\, \frac{\dd N^{\rm brem}}{\dd\, p_{Z'}^2 \dd \cos\theta_{Z'}^{}}\, \Theta(\Lambda_{\rm QCD}^{} - q^2) \cdot {\rm Acc}(\text{LLP},p_{Z'},\theta_{Z'})
\end{align}
where $N_p$ denotes the number of protons on target. We use the
FORESEE package~\cite{Kling:2021fwx} to compute the expected number of
signal events following the procedure outlined above. After implementing
the relevant production and decay expressions, we generate the
differential $Z^\prime$ spectra within FORESEE. We then derive projected
sensitivities to the general $U(1)$ gauge coupling as a function of
$M_{Z^\prime}$ for various charge assignments. The strongest sensitivities
are obtained in the region where the $Z^\prime$ decay length is comparable
to the distance between the FCC-hh interaction point and the FPF1(2)
detector.
\par Requiring
$N_{\rm event}^{\rm p\text{-}meson}
+N_{\rm event}^{\rm p\text{-}brem}>3$,
as obtained from Eqs.~(\ref{eq:num_p-meson1}) and
(\ref{eq:num_p-brem1}), corresponding to a $95\%$ confidence-level
sensitivity, we derive projected limits in the
$g_X$--$M_{Z^\prime}$ plane for different $U(1)_X$ charge assignments.
For $M_N > M_{Z^\prime}/3$, the projected sensitivities are identical
for the general $U(1)_X$ model (with $x_\Phi=1$) and the ``alternative'' scenario. Throughout this analysis, we consider the visible decay mode $Z^\prime\to$ visible, where ``visible'' denotes all charged final states. The projected sensitivities are shown in Fig.~\ref{params3} by the black dotted (FPF1($M_N>M_{Z'}/2$)) and dashed black (FPF2($M_N>M_{Z'}/2$)) curves, with each panel corresponding to a different choice of $U(1)_X$ charges. For FPF1, the sensitivity reaches $g_X\sim10^{-8}$ around $M_{Z^\prime}\sim\mathcal{O}(1)$ GeV, while the larger detector volume of FPF2 improves the reach by approximately one order of magnitude. Overall, FPF@FCC is expected to probe gauge couplings in the range $10^{-9}\lesssim g_X\lesssim5\times10^{-8}$ for $0.1~{\rm GeV}\lesssim M_{Z^\prime}\lesssim1.5~{\rm GeV}$, depending on the choice of $x_H$.

\par The upper-left panel of Fig.~\ref{params3} corresponds to the $U(1)_R$ model, obtained from the general $U(1)_X$ scenario with $x_H=-2$ and $x_\Phi=1$. Using the charge assignments in Tabs.~\ref{tab1} and \ref{tab2}, the same limits also apply to the $U(1)_{xq-\tau_R^3}$ model for $x=0$, although the $U(1)_R$ symmetry cannot be realized within the $U(1)_{q+\tilde{x}u}$ framework. The upper-right panel shows the case $x_H=-1$ and $x_\Phi=1$, for which the $e_R$ charge vanishes, modifying the branching ratio ${\rm BR}(Z^\prime\to\ell^+\ell^-)$. The same charge assignment can be obtained in the $U(1)_{xq-\tau_R^3}$ and $U(1)_{q+\tilde{x}u}$ scenarios by choosing $x=1/6$ and $\tilde{x}=-2$, respectively. The lower-left panel corresponds to the $B-L$ symmetry, which is reproduced by choosing $x_H=0$ and $x_\Phi=1$ in the general $U(1)_X$ model, $x=1/3$ in $U(1)_{xq-\tau_R^3}$, or $\tilde{x}=1$ in $U(1)_{q+\tilde{x}u}$. Finally, the lower-right panel shows the case $x_H=2$ and $x_\Phi=1$, which is equivalent to the choice $x=2/3$ in the $U(1)_{xq-\tau_R^3}$ model. This charge assignment cannot be realized within the $U(1)_{q+\tilde{x}u}$ framework, but it can be reproduced in the ``alternative'' scenario. Consequently, the same projected limits apply to both the general $U(1)_X$ and alternative scenarios. The same correspondence holds for all $Z^\prime$-mediated sensitivities presented in Figs.~\ref{params3}--\ref{params1}.

We compare our prospective limits with different existing exclusion bounds from electron beam-dump experiments involving Orsay \cite{Davier:1989wz}, KEK \cite{Beer:1986qr}, E141 \cite{Riordan:1987aw}, E137 \cite{Bjorken:1988as}, and E774 \cite{Bross:1989mp}. Existing bounds from different proton beam-dump experiments involving $\nu$-Cal \cite{Blumlein:2011mv,Blumlein:2013cua}, LSND \cite{LSND:1997vqj}, PS191 \cite{Bernardi:1985ny}, NOMAD \cite{NOMAD:2001eyx}, and CHARM \cite{CHARM:1985anb} are also shown. These  bounds are shown by shaded regions. We find that, in the mass range $0.1~{\rm GeV}\leq M_{Z^\prime}\leq1~{\rm GeV}$, the projected FPF1 sensitivity is comparable to the strongest existing constraints, while FPF2 surpasses the current experimental limits.
\subsubsection{Long-lived $Z^\prime$ and long-lived heavy neutrinos}
In this scenario, we consider a long-lived $Z^\prime$ produced through
meson decays and proton bremsstrahlung. Owing to its long lifetime, the
$Z^\prime$ can reach the FPF1(2) detector, where it decays either into
visible final states or into a pair of kinematically accessible heavy
neutrinos with $M_{N_{1,2}}<M_{Z^\prime}/2$, while the third heavy
neutrino satisfies $M_{N_3}>M_{Z^\prime}/2$, as illustrated in
Fig.~\ref{fig:llp1}. The produced heavy neutrinos are assumed to be
sufficiently long-lived to decay outside the FPF1(2) detector. This is
ensured by taking a small light--heavy neutrino mixing,
$|V_{\ell N}|^2=10^{-6}$. Consequently, the observable signal comes
entirely from the visible decay products of the $Z^\prime$. The expected
number of signal events is therefore obtained from
Eqs.~(\ref{eq:num_p-meson1}) and (\ref{eq:num_p-brem1}), with the only
modification being the reduced branching fraction
$\mathrm{BR}(Z^\prime\to\mathrm{visible})$ due to the additional decay
channel $Z^\prime\to NN$.
\par The resulting prospective limits are shown in Fig.~\ref{params3} for different general $U(1)$ charges by red dashed (general $U(1)$ scenarios)  and red dot-dashed (`alternative' (alt) scenario) for FPF1 detector. Whereas we show the expected limits by darker green dashed (general $U(1)$ scenarios) and darker green dot-dashed (`alternative' (alt) scenario) for FPF2 detector. We find that the prospective limits for the  `alternative' scenario is affected by the $U(1)_X$ charge of heavy neutrinos given in Tab.~\ref{tab2}. Since $\mathrm{BR}(Z^\prime\to NN)$ is larger in the ``alternative''
scenario than in the general $U(1)$ scenarios, the corresponding visible
branching fraction, $\mathrm{BR}(Z^\prime\to \mathrm{visible})$, is
suppressed in the alternative case. Consequently, the projected exclusion
regions in the $g_X$--$M_{Z^\prime}$ plane are generally broader for the
general $U(1)$ scenarios. The precise shape of the sensitivity contours,
however, depends on the $U(1)$ charge assignment, which affects both
$\mathrm{BR}(Z^\prime\to \mathrm{visible})$ and the couplings of the
$Z^\prime$ boson to SM quarks, thereby modifying the production rate.

For sufficiently small $M_{Z^\prime}$ and $g_X$, the sensitivities of the
general $U(1)$ scenarios (dashed curves) and the alternative scenario
(dot-dashed curves) become nearly identical for both FPF1 (red) and FPF2
(green). In this region, the exclusion reach exhibits only a weak
dependence on $\mathrm{BR}(Z^\prime\to NN)$. Overall, our results indicate
that the FPF@FCC has the potential to substantially improve the current
constraints on the $g_X$--$M_{Z^\prime}$ parameter space.
\subsubsection{Short-lived $Z^\prime$ and long-lived heavy neutrinos}
\begin{figure*}[htb!]
\centering
\includegraphics[scale=0.43]{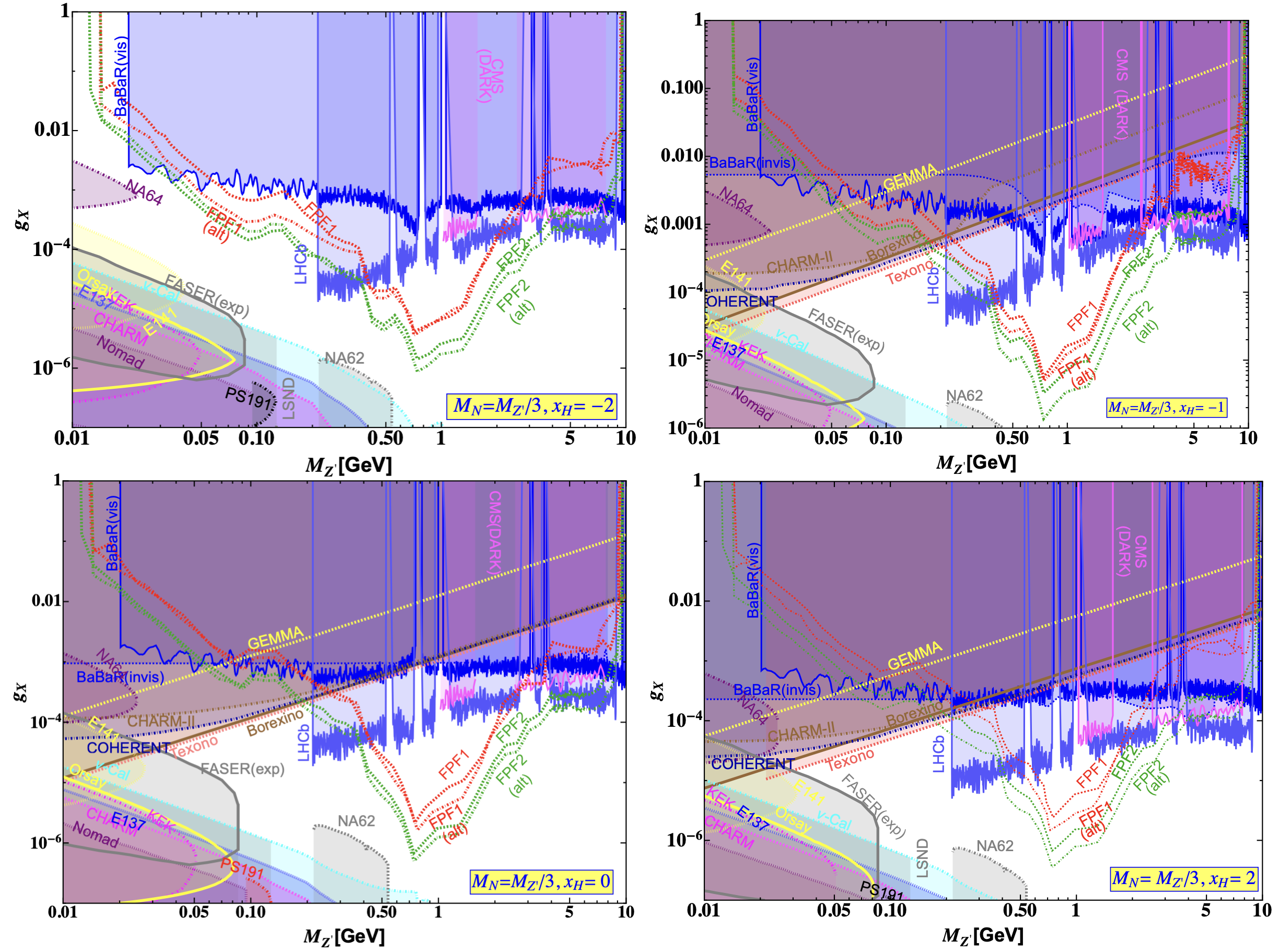}
\caption{Prospective limits on the general $U(1)$ gauge coupling as a function of $M_{Z^\prime}$ for a short-lived $Z^\prime$ boson producing long-lived heavy neutrinos with $M_{N_{1,2}}=M_{Z^\prime}/3$. The dashed red (dark-green) curves correspond to the general $U(1)$ scenario for FPF1 (FPF2), while the red (dark-green) dot-dashed curves denote the
corresponding alternative (\textit{alt}) scenario. In all cases, we assume $|V_{\ell N}|^2=10^{-6}$. The shaded regions are excluded by
existing experimental constraints.}
\label{params4}
\end{figure*}
In this scenario, we consider a short-lived $Z^\prime$ produced through meson decays and bremsstrahlung, which promptly decays at the FCC-hh interaction point into a pair of kinematically accessible long-lived heavy neutrinos satisfying $M_N<M_{Z^\prime}/2$. Owing to the magnetic shielding between the interaction point and the FPF1(2) detector, the Standard Model particles produced in the $Z^\prime$ decay do not reach the detector. In contrast, the long-lived heavy neutrinos travel to the FPF1(2) detector and subsequently decay into visible final states, as illustrated in Fig.~\ref{fig:llp1}. Consequently, the observable signal arises entirely from the visible decays of the heavy neutrinos. The corresponding signal acceptance is given by
\begin{align}
\label{eq:accep_2}
\text{Acc}({\rm LLP}, p_i^1, \theta_i^1, p_i^2, \theta_i^2)
=
\text{BR}(Z^\prime \to 2N_i)
\Big[
\text{Acc}(N_i^1,p_i^1,\theta_i^1)
+
\text{Acc}(N_i^2,p_i^2,\theta_i^2)
\Big],
\end{align}
where $\vec{p}_{Z^\prime}=\vec{p}_i^{\,1}+\vec{p}_i^{\,2}$ and $N_i^{1(2)}$ denotes the first (second) heavy neutrino of generation $i$ produced in the decay $Z^\prime\to N_iN_i$ with momentum $p_i^{1(2)}$. The acceptance for each heavy neutrino is
\begin{align}
\text{Acc}(N_i^\alpha,p_i^\alpha,\theta_i^\alpha)
=
\mathcal{P}(p_i^\alpha,\theta_i^\alpha)\,
\text{BR}(N_i\to{\rm visible})~.
\end{align}

This signature is relevant only when the $Z^\prime$ is sufficiently short-lived, corresponding to relatively large gauge couplings and/or a large $Z^\prime$ mass. In the high-mass regime, bremsstrahlung dominates the $Z^\prime$ production. Since the $Z^\prime$ decays into two identical, highly boosted heavy neutrinos in the forward direction, one has $p_{N_i}\simeq p_{Z^\prime}/2$ and $\theta_{N_i}\simeq\theta_{Z^\prime}$. Consequently, the momentum and angular distributions of the heavy neutrinos closely follow those of the parent $Z^\prime$, allowing the acceptance to be approximated as
\bea
\label{eq:accep_22}
\text{Acc}({\rm LLP}, p_i^1,\theta_i^1,p_i^2,\theta_i^2)
&\equiv&
\text{Acc}({\rm LLP},p_{N_i},\theta_{N_i})\nonumber\\
&\approx&
2\,\text{BR}(Z^\prime\to2N_i)\,
\text{Acc}(N_i,p_{N_i},\theta_{N_i}),
\eea
\begin{figure*}[h]
\centering
\includegraphics[scale=0.42]{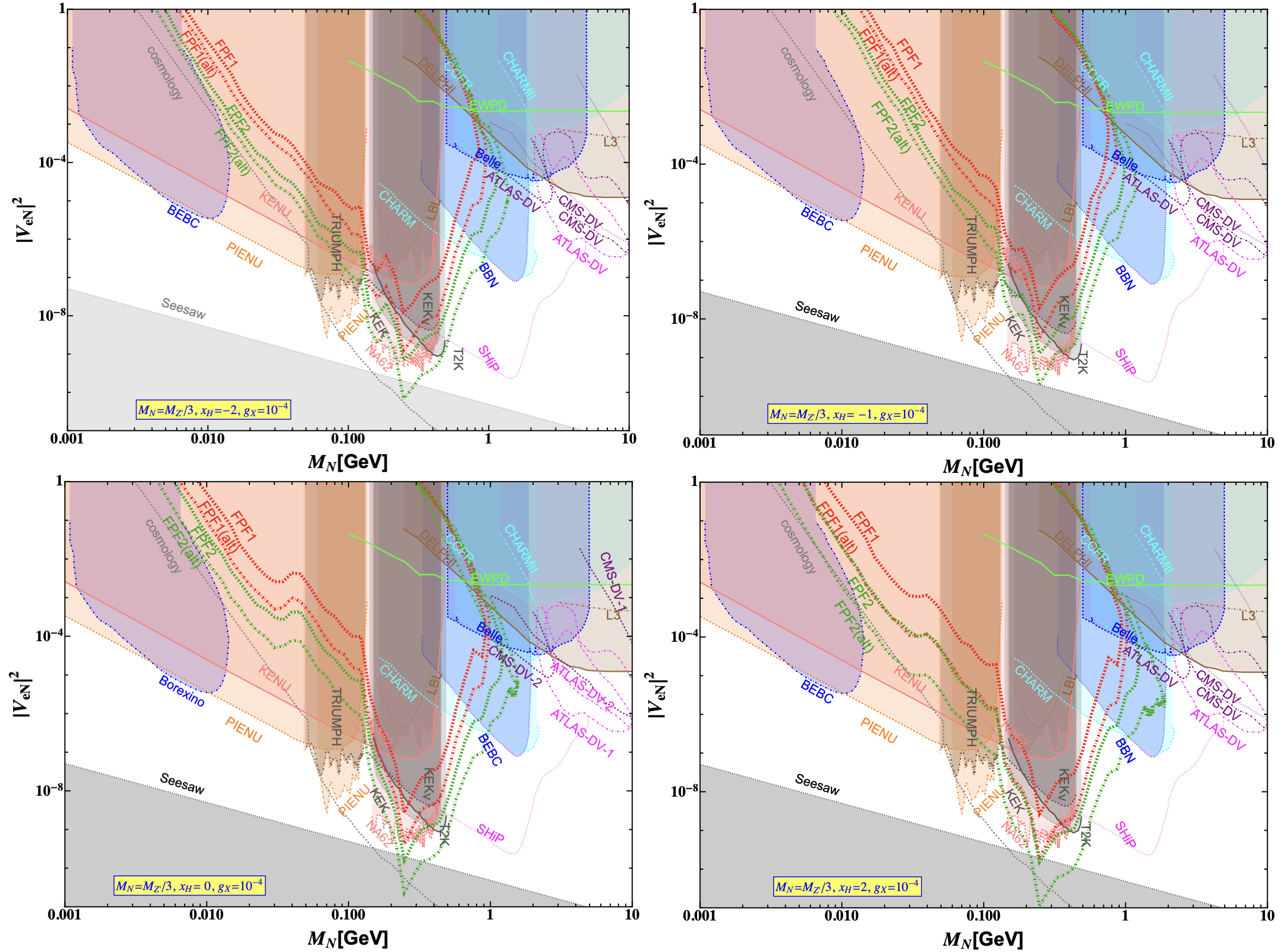}
\caption{Prospective limits on $|V_{eN}|^2$ as a function of $M_N$ for long-lived
heavy neutrinos produced via the decay of a short-lived $Z^\prime$, assuming $g_X=10^{-4}$ and $M_{N_{1,2}}=M_{Z^\prime}/3$. The dashed red
(dark-green) curves correspond to the general $U(1)_X$ scenario, while the dot-dashed red (dark-green) curves denote the corresponding
alternative (\textit{alt}) scenario, for FPF1 (FPF2). The shaded regions are excluded by existing experimental constraints and theoretical
estimations.}
\label{params1}
\end{figure*}
\begin{figure*}[h]
\centering
\includegraphics[scale=0.42]{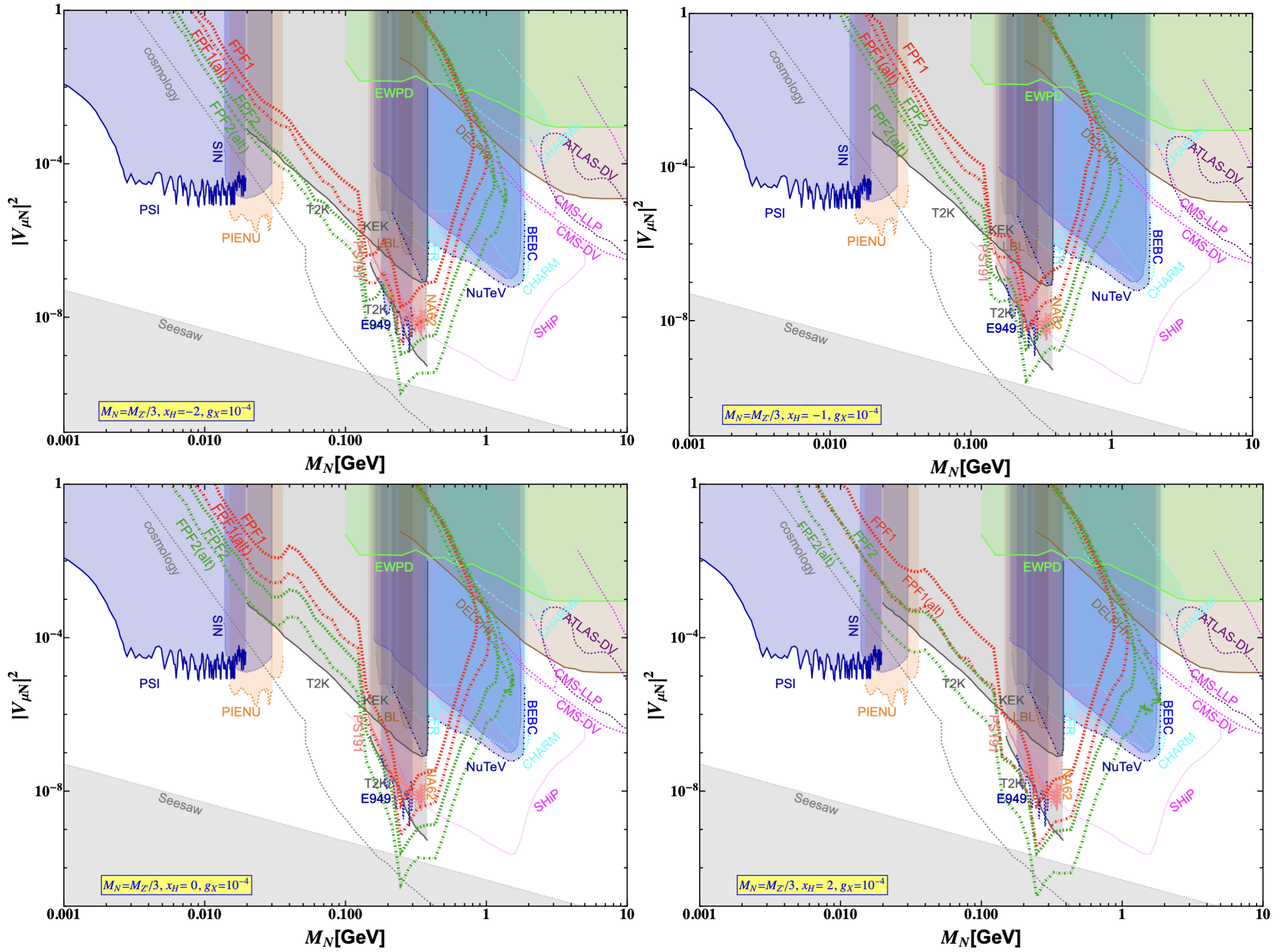}
\caption{Same as Fig.~\ref{params1} but for mixing $|V_{\mu N}|^2$. See text for details.}
\label{params2}
\end{figure*}
where $p_{N_i}\simeq p_{Z^\prime}/2$ and $\theta_{N_i}\simeq\theta_{Z^\prime}$. Using this approximation, the expected numbers of signal events from the bremsstrahlung and meson decay channels are given by
\begin{align}
\label{eq:num_p-brem3}
N_{\rm event}^{\rm p\mathchar`-brem}
&=
N_p\, |F_1(m_{Z^\prime}^2)|^2
\int \dd p_{Z^\prime}^2
\int \dd\cos\theta_{Z^\prime}
\frac{\dd N^{\rm brem}}
{\dd p_{Z^\prime}^2\,\dd\cos\theta_{Z^\prime}}
\Theta(\Lambda_{\rm QCD}-q^2)
\cdot
\sum_i
{\rm Acc}({\rm LLP},p_{N_i},\theta_{N_i}),
\\
\label{eq:num_p-meson3}
N_{\rm event}^{\rm p\mathchar`-meson}
&=
N_p
\sum_{M=\pi^0,\eta}
\int \dd p_M^2
\int \dd\cos\theta_M
\int \dd p_{Z^\prime}^2
\int \dd\cos\theta_{Z^\prime}
\frac{\dd N^M}
{\dd p_M^2\,\dd\cos\theta_M}
\cdot
\sum_i
{\rm Acc}({\rm LLP},p_{N_i},\theta_{N_i}),
\end{align}
respectively, where $N_p$ denotes the number of protons on target. This search channel is particularly sensitive to scenarios with relatively large $U(1)$ gauge couplings and/or heavy $Z^\prime$ masses, while simultaneously probing small light-heavy neutrino mixing, $|V_{\ell N}|^2$, and low heavy-neutrino masses at FPF@FCC.
\par Assuming a $95\%$ confidence-level sensitivity corresponding to
$N_{\rm event}^{\rm p\mathchar`-meson}
+N_{\rm event}^{\rm p\mathchar`-brem}>3$,
with the event yields obtained from
Eqs.~(\ref{eq:num_p-brem3}) and (\ref{eq:num_p-meson3}),
we derive projected limits in the $g_X$--$M_{Z^\prime}$ plane for
various $U(1)_X$ charge assignments, fixing
$|V_{\ell N}|^2=10^{-6}$. In this scenario, the signal originates from the visible decays of heavy neutrinos through $N\to\ell/\nu_\ell+$ associated particles. The projected sensitivities for the general $U(1)$ and ``alternative'' scenarios are shown by the dashed red (FPF1) and dot-dashed red (FPF1(alt)) curves, respectively, while the corresponding FPF2 projections are represented by the dashed green and dot-dashed green curves in Fig.~\ref{params4}. We find that FPF1 can probe gauge couplings down to $g_X\sim10^{-6}$ around $M_{Z^\prime}\simeq0.75$ GeV, with the sensitivity improving for larger values of $x_H$. Owing to its larger detector volume, FPF2 further enhances the reach by approximately one order of magnitude.
\par We compare our projections with the existing constraints from neutrino-electron scattering experiments, including TEXONO \cite{TEXONO:2009knm}, BOREXINO \cite{Borexino:2008gab,Bellini:2011rx}, GEMMA \cite{Beda:2010hk}, and CHARM-II \cite{CHARM-II:1994dzw}, as well as neutrino-nucleon scattering measurements from COHERENT \cite{COHERENT:2020ybo} and dark photon searches at LHCb \cite{LHCb:2019vmc}, BaBar \cite{BaBar:2014zli,BaBar:2017tiz}, and CMS~\cite{CMS:2023slr}. The shaded regions in Fig.~\ref{params4} are excluded by these experiments. In the $U(1)_R$ scenario, the absence of $Z^\prime$-mediated neutrino-electron scattering removes the corresponding scattering constraints, thereby enlarging the parameter space accessible at FPF@FCC. We find that, independent of the choice of $x_H$, FPF@FCC improves upon the existing limits in the range $0.4~{\rm GeV}\lesssim M_{Z^\prime}\lesssim4~{\rm GeV}$, while in the $U(1)_R$ scenario the sensitivity extends to $M_{Z^\prime}\lesssim0.2~{\rm GeV}$, as shown in the upper-left panel of Fig.~\ref{params4}. Finally, we emphasize that the projected limits in Fig.~\ref{params4} depend on the assumed light-heavy neutrino mixing, $|V_{\ell N}|^2$, which determines the probability for the heavy neutrinos to decay inside the detector and hence directly affects the signal yield.
\par From the short-lived $Z^\prime$ and long-lived heavy neutrino scenarios, one can also estimate bounds on light-heavy neutrino mixing using Eqs.~(\ref{eq:num_p-brem3}) and (\ref{eq:num_p-meson3}). In general, for a fixed value of $x_H$, the signal yield depends on four free parameters: $g_X$, $M_{Z^\prime}$, $M_N$, and $|V_{\ell N}|^2$. Therefore, to study the sensitivity in the $M_N$--$|V_{\ell N}|^2$ plane, the values of $g_X$ and $M_{Z^\prime}$ must be fixed. For the short-lived $Z^\prime$ scenario, we adopt the benchmark choice $g_X=10^{-4}$ and fix the heavy neutrino mass to be $M_N=M_{Z^\prime}/3$. We show the prospective bounds on $|V_{e N}|^2$ and $|V_{\mu N}|^2$ as a function of $M_N$ in Figs.~\ref{params1} and \ref{params2}, respectively. We show the prospective bounds from general $U(1)$ scenario by dashed red (darker green) and `alternative' (alt) scenario by  dot dashed red (darker green) line for FPF1(2) detector in FPF@FCC experiment. We find that heavy-neutrino pair production can probe the light--heavy neutrino mixing down to
$|V_{(e,\mu)N}|^2 \sim 10^{-9}$ for
$M_N \simeq 0.25~\mathrm{GeV}$, depending on the
$U(1)$ charge assignment and the FPF detector configuration. More generally, this channel is sensitive to $10^{-9}\lesssim |V_{(e,\mu)N}|^2 \lesssim 10^{-6}$ for $0.25~\mathrm{GeV}\lesssim M_N \lesssim 0.85~\mathrm{GeV}$, with the precise reach depending on the final state, the $U(1)$ charge assignment, and the detector geometry.
\section{Conclusions}
\label{sec:conc}
In this work, we have investigated the prospects for probing long-lived heavy neutrinos and light gauge bosons at the proposed Forward Physics Facility~(FPF) of the FCC-hh within a broad class of anomaly-free chiral $U(1)$ extensions of the SM. We considered both the minimal realization with universal RHN charges and an alternative scenario with non-universal charge assignments, encompassing various well-known models. In these frameworks, the heavy neutrinos generate light neutrino masses through the type-I seesaw mechanism, interact with the SM weak gauge bosons via the light-heavy neutrino mixing, and couple directly to the new gauge boson $Z^\prime$ through the additional $U(1)$ gauge interaction.
\par We analyzed four complementary long-lived particle signatures at the FPF. First, we studied heavy neutrinos produced through meson decays and derived the projected sensitivities in the $(|V_{\ell N}|^2,M_N)$ plane using their visible decays. We then investigated three $Z^\prime$-mediated scenarios: (i) a long-lived $Z^\prime$ decaying into visible final states when the
channel $Z^\prime\to NN$ is kinematically inaccessible; (ii) a long-lived $Z^\prime$ decaying into both visible SM final states
and a pair of heavy neutrinos, where the heavy neutrinos are sufficiently long-lived to escape the detector before decaying; and (iii) a promptly decaying $Z^\prime$ producing long-lived heavy neutrinos that subsequently decay inside the detector volume. These scenarios allow us to probe both the $(g_X,M_{Z^\prime})$ and $(|V_{\ell N}|^2,M_N)$ parameter spaces through complementary visible signatures.
\par For heavy neutrinos produced from meson decays, we find that FPF1 achieves a sensitivity comparable to the projected SHiP reach and even outperforms SHiP for $2.5~{\rm GeV}\lesssim M_N\lesssim6~{\rm GeV}$ owing to the significantly enhanced $B$-meson production at the FCC-hh. Benefiting from its much larger decay volume, FPF2 substantially extends the accessible parameter space, surpassing both current experimental limits and the projected sensitivities of future experiments over a broad region of the $(|V_{\ell N}|^2,M_N)$ plane. For the $Z^\prime$-mediated channels, we show that FPF@FCC can probe gauge couplings as small as $g_X\sim10^{-9}$ for sub-GeV $Z^\prime$ masses, depending on the underlying $U(1)$ charge assignment. While the scenario with a long-lived $Z^\prime$ decaying into heavy-neutrino pairs weakens the sensitivity because of the reduced visible branching fraction, the prompt-$Z^\prime$ scenario provides strong and complementary constraints through the visible decays of long-lived heavy neutrinos.
\par Overall, our results demonstrate that the FPF at the FCC-hh offers an exceptional opportunity to explore light long-lived heavy neutrinos and light $Z^\prime$ bosons predicted in the anomaly-free $U(1)$ extensions of the SM. The combination of enormous forward particle flux, enhanced
heavy-flavor production, and the large fiducial decay volume—particularly in the FPF2 configuration—provides sensitivity well beyond existing and projected experiments over broad regions of
the model parameter space. The FPF@FCC therefore constitutes a powerful and complementary facility for testing the origin of neutrino masses and searching for hidden gauge sectors through long-lived particle signatures.
\black
\begin{acknowledgements}
S.K.A is supported by JST SPRING, Grant Number JPMJSP2119. The work of S.M. is supported by KIAS Individual Grants (PG086002) at Korea Institute for Advanced Study. 
\end{acknowledgements}
\appendix
\section{Decay widths of heavy neutrinos}
\label{appl:decay_width}
In this section we summarize the heavy neutrino decay channels relevant to our analysis. For $M_N<m_W$, heavy neutrinos decay through off-shell $W$ and $Z$ bosons into leptonic and semileptonic final states. The corresponding leptonic partial decay widths are given below.
For the charged-current process mediated by an off-shell $W$ boson,
\bea
\Gamma(N\to  \ell_1^- \ell_2^+ \nu_{\ell_2})=
|V_{\ell_1 N}|^2
\frac{G_F^2}{16\pi^3}
M_N^5
(1-\delta_{\ell_1\ell_2})
I_1(y_{\nu_{\ell_2}},y_{\ell_1},y_{\ell_2}),
\label{nul1l2}
\eea
where $G_F$ is the Fermi constant, $\delta_{\ell_1\ell_2}$ is the Kronecker delta, and $y_i\equiv m_i/M_N$.
The neutral-current decay through an off-shell $Z$ boson into charged leptons is given by
\begin{align}
\Gamma(N\to \nu_{\ell_1}\ell_2^-\ell_2^+)
=&
|V_{\ell_1N}|^2
\frac{G_F^2}{8\pi^3}
M_N^5
\Big[
2(g_L^\ell g_R^\ell+g_R^\ell\delta_{\ell_1\ell_2})
I_2(y_{\nu_{\ell_1}},y_{\ell_2},y_{\ell_2})
\nonumber\\
&
+\Big((g_L^\ell)^2+(g_R^\ell)^2
+(1+2g_L^\ell)\delta_{\ell_1\ell_2}\Big)
I_1(y_{\nu_{\ell_1}},y_{\ell_2},y_{\ell_2})
\Big],
\label{nuiljlj}
\end{align}
where
$g_L^\ell=-1/2+\sin^2\theta_W$
and
$g_R^\ell=\sin^2\theta_W$.
Equation~\eqref{nuiljlj} includes both
$N\to\nu_{\ell_1}\ell_2^-\ell_2^+$ and
$N\to\bar{\nu}_{\ell_1}\ell_2^-\ell_2^+$.
The invisible decay mode into three neutrinos is
\bea
\sum_{\ell_2=e,\mu,\tau}
\Gamma(N\to\nu_{\ell_1}\nu_{\ell_2}\bar{\nu}_{\ell_2})
=
|V_{\ell_1N}|^2
\frac{G_F^2}{96\pi^3}
M_N^5,
\label{3nu}
\eea
where the corresponding charge-conjugate process has also been included. The light neutrino masses are neglected in this calculation.
The semileptonic decay widths into pseudoscalar mesons are
\begin{align}
\Gamma(N\to \ell_1^-P^+)
&=
|V_{\ell_1N}|^2
\frac{G_F^2}{16\pi}
M_N^3
f_P^2
|V_P|^2
F_P(y_\ell,y_P),
\label{l1p}\\
\Gamma(N\to\nu_{\ell_1}P^0)
&=
|V_{\ell_1N}|^2
\frac{G_F^2}{2\pi}
M_N^3
f_P^2
\kappa_P^2
F_P(y_{\nu_\ell},y_P),
\label{nup}
\end{align}
where $f_P$ is the pseudoscalar decay constant, $V_P$ denotes the corresponding CKM matrix element, and $\kappa_P$ is the neutral-current coupling,
\begin{align}
\kappa_{\pi^0}
=
-\frac{1}{2\sqrt2},
\qquad
\kappa_\eta
=
-\frac{1}{2\sqrt6},
\qquad
\kappa_{K^0}
=
\frac14.
\end{align}

Similarly, the decay widths into vector mesons are
\begin{align}
\Gamma(N\to \ell_1^-V^+)
&=
|V_{\ell_1N}|^2
\frac{G_F^2}{16\pi}
M_N^3
f_V^2
|V_V|^2
F_V(y_\ell,y_V),
\label{l1v}\\
\Gamma(N\to\nu_{\ell_1}V^0)
&=
|V_{\ell_1N}|^2
\frac{G_F^2}{2\pi}
M_N^3
f_V^2
\kappa_V^2
F_V(y_{\nu_\ell},y_V),
\label{nuv}
\end{align}
where $f_V$ denotes the vector meson decay constant and
\begin{align}
\kappa_{\rho^0}
=
\frac{1}{\sqrt2}
\left(\frac12-\sin^2\theta_W\right),\,\,\,
\kappa_\omega
=
-\frac{\sin^2\theta_W}{3\sqrt2},\,\,\,
\kappa_{K^{*0}}
=
\frac12
\left(
\frac23\sin^2\theta_W-\frac12
\right).
\end{align}
For $M_N>\mu_0$, the hadronic final states are described in terms of quark degrees of freedom. The charged-current decay width is
\bea
\Gamma(N\to\ell_1^-q\bar{q}')
=
N_C
|V_{\ell_1N}|^2
|V_{q\bar{q}'}|^2
\frac{G_F^2}{16\pi^3}
M_N^5
I_1(y_{\nu_{\ell_1}},y_q,y_{q'}),
\label{l1ud}
\eea
while the neutral-current decay width is
\bea
\Gamma(N\to\nu_{\ell_1}q\bar{q})
=
N_C
|V_{\ell_1N}|^2
\frac{G_F^2}{8\pi^3}
M_N^5
\Big[
2g_L^qg_R^qI_2
+
\big((g_L^q)^2+(g_R^q)^2\big)I_1
\Big],
\label{nuqq}
\eea
where the arguments of $I_{1,2}$ are
$(y_{\nu_\ell},y_q,y_q)$.
The neutral-current couplings are
\begin{align}
g_L^u
&=
\frac12-\frac23\sin^2\theta_W,
&
g_R^u
&=
-\frac23\sin^2\theta_W,\\
g_L^d
&=
-\frac12+\frac13\sin^2\theta_W,
&
g_R^d
&=
\frac13\sin^2\theta_W.
\end{align}
\begin{table}[]
    \centering
    \begin{tabular}{|c|c|c|c||c|c|c|c|}\hline
       Psudoscalar(P)  & $m_P$ (MeV) & $f_P$(MeV) & $V_P$ & Vector(V)  & $m_V$ (MeV) & $f_V$(MeV) & $V_V$ \\ \hline
        $\pi^\pm$ & 139.6 & 130.7 & $V_{ud}$ & $\rho^\pm$ & 775.8 & 220 & $V_{ud}$ \\ \hline
        $K^\pm$ & 493.7 & 159.8 & $V_{us}$ & $K^{*\pm}$ & 891.66 & 217 & $V_{us}$ \\ \hline
        $\eta$ & 547.8 & 164.7 & -- & $\omega$ & 782.59 & 195 & -- \\ \hline
        $\pi^0$ & 135 & 130 & -- & $\rho^0$ & 776 & 220 & -- \\ \hline
        $K^0$ & 497.6 & 159 & -- & $K^{*0}$ & 896.1 & 217  & -- \\ \hline
    \end{tabular}
    \caption{Masses and decay constants of pseudoscalar and vector mesons following \cite{ParticleDataGroup:2010dbb,CLEO:2005jsh,MILC:2002lnl,Ivanov:2006ni,Feldmann:1999uf,Cvetic:2004qg,Ebert:2006hj}.}
    \label{tab:mesdec}
\end{table}
The masses, decay constants, and CKM matrix elements used for the pseudoscalar and vector mesons are summarized in Tab.~\ref{tab:mesdec}. The kinematic functions appearing in the above expressions are
\begin{align}
I_1(x,y,z)
&=
\int_{(x+y)^2}^{(1-z)^2}
\frac{ds}{s}
(s-x^2-y^2)
(1+z^2-s)
\lambda(s,x^2,y^2)
\lambda(1,s,z^2),
\\
I_2(x,y,z)
&=
yz
\int_{(y+z)^2}^{(1-x)^2}
\frac{ds}{s}
(1+x^2-s)
\lambda(s,y^2,z^2)
\lambda(1,s,x^2),
\\
F_P(x,y)
&=
\lambda(1,x^2,y^2)
\Big[(1+x^2)(1+x^2-y^2)-4x^2\Big],
\\
F_V(x,y)
&=
\lambda(1,x^2,y^2)
\Big[(1-x^2)^2+(1+x^2)y^2-2y^4\Big],
\end{align}
where the K\"all\'en function is
\begin{equation}
\lambda(x,y,z)
=
\sqrt{x^2+y^2+z^2-2xy-2yz-2zx}.
\end{equation}

Since the heavy neutrinos are Majorana fermions in the general $U(1)$ models discussed in Sec.~\ref{sec:model}, all charged-current decay modes receive equal contributions from their charge-conjugate processes, doubling the corresponding partial widths. This enhancement is absent for Dirac neutrinos.
The total decay width is obtained by summing all leptonic and semileptonic channels,
\small
\begin{equation}
\Gamma_N
=
\sum_{\ell_1,\ell_2(\ell_1\neq\ell_2)}
\left[
2\Gamma(N\to\ell_1^-\ell_2^+\nu_{\ell_2})
+
\Gamma(N\to\nu_{\ell_1}\ell_2^-\ell_2^+)
\right]
+
\sum_{\ell_2}
\Gamma(N\to\nu_{\ell_2}\ell_2^-\ell_2^+)
+
\sum_{\ell_1}
\Gamma(N\to\nu_{\ell_1}\nu\bar{\nu})
+
\Gamma^{\rm semilepton},
\label{decn}
\end{equation}
\normalsize
where the semileptonic contribution is
\begin{equation}
\begin{aligned}
\Gamma^{\rm semilepton}
=
&
\theta(\mu_0-M_N)
\sum_{\ell_1,P,V}
\Big[
2\Gamma(N\to\ell_1^-P^+)
+
2\Gamma(N\to\ell_1^-V^+)
+
\Gamma(N\to\nu_{\ell_1}P^0)
+
\Gamma(N\to\nu_{\ell_1}V^0)
\Big]
\\
&
+\theta(M_N-\mu_0)
\sum_{\ell_1,q,q'}
\Big[
\Gamma(N\to\nu_{\ell_1}q\bar{q})
+
2\Gamma(N\to\ell_1^-q\bar{q}')
\Big].
\end{aligned}
\label{seml}
\end{equation}
The visible decay width includes all channels except the invisible mode $N\to3\nu$. Throughout this work we take $\mu_0=957.8$ MeV as the transition scale between the exclusive meson description and the inclusive quark-level treatment of semileptonic decays. 
\bibliographystyle{utphys}
\bibliography{bibliography}
\end{document}